\documentclass[11pt]{article}

\usepackage[final]{acl}

\usepackage{times}
\usepackage{latexsym}
\usepackage{fvextra}
\usepackage{booktabs}
\usepackage{amsmath}
\usepackage{adjustbox}
\usepackage{amssymb} 
\usepackage{enumitem}
\usepackage{contour}
\usepackage{multirow}
\contourlength{0.5pt}
\usepackage{xcolor}
\usepackage{pgfplots}
\pgfplotsset{compat=1.18}
\usepgfplotslibrary{groupplots}
\usetikzlibrary{plotmarks}
\pgfdeclareplotmark{halfsquareh}{%
    \pgfscope%
    \pgfsetfillcolor{white}%
    \pgfpathmoveto{\pgfqpoint{-\pgfplotmarksize}{0pt}}%
    \pgfpathlineto{\pgfqpoint{\pgfplotmarksize}{0pt}}%
    \pgfpathlineto{\pgfqpoint{\pgfplotmarksize}{\pgfplotmarksize}}%
    \pgfpathlineto{\pgfqpoint{-\pgfplotmarksize}{\pgfplotmarksize}}%
    \pgfpathclose%
    \pgfusepath{fill}%
    \endpgfscope%
    \pgfpathmoveto{\pgfqpoint{-\pgfplotmarksize}{-\pgfplotmarksize}}%
    \pgfpathlineto{\pgfqpoint{\pgfplotmarksize}{-\pgfplotmarksize}}%
    \pgfpathlineto{\pgfqpoint{\pgfplotmarksize}{0pt}}%
    \pgfpathlineto{\pgfqpoint{-\pgfplotmarksize}{0pt}}%
    \pgfpathclose%
    \pgfusepath{fill}%
    \pgfpathmoveto{\pgfqpoint{-\pgfplotmarksize}{-\pgfplotmarksize}}%
    \pgfpathlineto{\pgfqpoint{\pgfplotmarksize}{-\pgfplotmarksize}}%
    \pgfpathlineto{\pgfqpoint{\pgfplotmarksize}{\pgfplotmarksize}}%
    \pgfpathlineto{\pgfqpoint{-\pgfplotmarksize}{\pgfplotmarksize}}%
    \pgfpathclose%
    \pgfusepath{stroke}%
}
\pgfdeclareplotmark{stripedsquare}{%
    \pgfpathrectangle
        {\pgfqpoint{-\pgfplotmarksize}{-\pgfplotmarksize}}
        {\pgfqpoint{2\pgfplotmarksize}{2\pgfplotmarksize}}
    \pgfusepath{fill,stroke}%
    \pgfsetlinewidth{0.5pt}
    \pgfpathmoveto{\pgfqpoint{-\pgfplotmarksize}{0.33\pgfplotmarksize}}
    \pgfpathlineto{\pgfqpoint{\pgfplotmarksize}{0.33\pgfplotmarksize}}
    \pgfpathmoveto{\pgfqpoint{-\pgfplotmarksize}{-0.33\pgfplotmarksize}}
    \pgfpathlineto{\pgfqpoint{\pgfplotmarksize}{-0.33\pgfplotmarksize}}
    \pgfusepath{stroke}%
}
\pgfdeclareplotmark{stripeddiamond}{%
    \pgfpathmoveto{\pgfqpoint{0pt}{\pgfplotmarksize}}
    \pgfpathlineto{\pgfqpoint{\pgfplotmarksize}{0pt}}
    \pgfpathlineto{\pgfqpoint{0pt}{-\pgfplotmarksize}}
    \pgfpathlineto{\pgfqpoint{-\pgfplotmarksize}{0pt}}
    \pgfpathclose
    \pgfusepath{fill,stroke}%
    \pgfsetlinewidth{0.5pt}
    \pgfpathmoveto{\pgfqpoint{-0.67\pgfplotmarksize}{0.33\pgfplotmarksize}}
    \pgfpathlineto{\pgfqpoint{0.67\pgfplotmarksize}{0.33\pgfplotmarksize}}
    \pgfpathmoveto{\pgfqpoint{-0.67\pgfplotmarksize}{-0.33\pgfplotmarksize}}
    \pgfpathlineto{\pgfqpoint{0.67\pgfplotmarksize}{-0.33\pgfplotmarksize}}
    \pgfusepath{stroke}%
}
\pgfdeclareplotmark{diamondplain}{%
    \pgfpathmoveto{\pgfqpoint{0pt}{\pgfplotmarksize}}
    \pgfpathlineto{\pgfqpoint{\pgfplotmarksize}{0pt}}
    \pgfpathlineto{\pgfqpoint{0pt}{-\pgfplotmarksize}}
    \pgfpathlineto{\pgfqpoint{-\pgfplotmarksize}{0pt}}
    \pgfpathclose
    \pgfusepath{fill,stroke}%
}
\pgfdeclareplotmark{stripedcircle}{%
    \pgfpathcircle{\pgfpointorigin}{\pgfplotmarksize}
    \pgfusepath{fill,stroke}%
    \pgfsetlinewidth{0.5pt}
    \pgfpathmoveto{\pgfqpoint{-0.944\pgfplotmarksize}{0.33\pgfplotmarksize}}
    \pgfpathlineto{\pgfqpoint{0.944\pgfplotmarksize}{0.33\pgfplotmarksize}}
    \pgfpathmoveto{\pgfqpoint{-0.944\pgfplotmarksize}{-0.33\pgfplotmarksize}}
    \pgfpathlineto{\pgfqpoint{0.944\pgfplotmarksize}{-0.33\pgfplotmarksize}}
    \pgfusepath{stroke}%
}
\pgfdeclareplotmark{stripedtriangle}{%
    \pgfpathmoveto{\pgfqpoint{0pt}{\pgfplotmarksize}}
    \pgfpathlineto{\pgfqpoint{-\pgfplotmarksize}{-\pgfplotmarksize}}
    \pgfpathlineto{\pgfqpoint{\pgfplotmarksize}{-\pgfplotmarksize}}
    \pgfpathclose
    \pgfusepath{fill,stroke}%
    \pgfsetlinewidth{0.5pt}
    \pgfpathmoveto{\pgfqpoint{-0.335\pgfplotmarksize}{0.33\pgfplotmarksize}}
    \pgfpathlineto{\pgfqpoint{0.335\pgfplotmarksize}{0.33\pgfplotmarksize}}
    \pgfpathmoveto{\pgfqpoint{-0.665\pgfplotmarksize}{-0.33\pgfplotmarksize}}
    \pgfpathlineto{\pgfqpoint{0.665\pgfplotmarksize}{-0.33\pgfplotmarksize}}
    \pgfusepath{stroke}%
}
\usepackage{subcaption}
\usepackage[T1]{fontenc}

\usepackage[utf8]{inputenc}

\usepackage{microtype}

\usepackage{inconsolata}

\usepackage{graphicx}

\usepackage{makecell}
\usepackage{siunitx}

\newcommand{\lat}[4][1]{%
  \makecell[c]{%
    \textbf{\num[round-mode=places,
      round-precision=#1]{#2}}\\[-1pt]
    {\scriptsize
      [\num[round-mode=places,round-precision=#1]{#3},
       \num[round-mode=places,round-precision=#1]{#4}]}%
  }%
}

\definecolor{frontiercolor}{RGB}{213,94,0}

\definecolor{mutedblue}{rgb}{0.54, 0.81, 0.94}
\definecolor{mutedorange}{RGB}{221,132,82}
\definecolor{mutedgreen}{RGB}{83, 246, 165}
\definecolor{mutedred}{RGB}{252, 165, 165}
\definecolor{mutedpurple}{RGB}{129,114,179}
\definecolor{mutedgold}{RGB}{207,177,92}
\definecolor{mutedteal}{RGB}{76,158,154}
\definecolor{typesafe}{RGB}{228,81,186}

\title{Decision-Oriented Recommendation Reranking: An Empirical Study of Its Promise and Limitations}

\author{Hanjia Lyu \\
  Singapore Management University \\
  \texttt{hjlyu@smu.edu.sg} \\\And
  Yinglong Xia \\
  Meta AI\\
  \texttt{yxia@meta.com} \\}

\begin{document}
\maketitle
\begin{abstract}
Large language models (LLMs) have shown promise for recommendation reranking, but their use introduces an important trade-off between recommendation quality and serving efficiency. We investigate whether decision-oriented systems provide a useful alternative when the reranking task is fundamentally a structured choice among predefined candidate items. 
Specifically, we conduct a controlled empirical study of two decision-oriented systems, TypeSafe Jev and OpenAI Decisions, for recommendation reranking. Across three Amazon Reviews domains and four candidate-set sizes, we compare these systems against recommendation-specific models and pointwise and listwise LLM-based rerankers, systematically evaluating recommendation effectiveness, candidate-order robustness, and efficiency and scalability. 
Our results reveal that Jev frequently achieves a favorable quality--latency trade-off, offering competitive recommendation effectiveness at substantially lower observed latency than stronger LLM-based alternatives. 
Overall, our findings suggest that specialization for decision-making does not inherently guarantee superior recommendation quality or robustness. Instead, the practical value of decision-oriented inference depends on implementation-specific characteristics and application requirements.
These findings motivate further investigation of decision-oriented systems for recommendation and other ranking tasks with structured output spaces.
\end{abstract}

\section{Introduction}
\label{sec:intro}

Recommender systems commonly adopt multi-stage architectures in which an efficient retrieval model first identifies a manageable set of candidate items and a more expressive ranking model subsequently determines their final ordering~\cite{covington2016deep}. 
This separation allows the ranking stage to leverage richer representations and more computationally intensive models that are typically infeasible during large-scale retrieval.
Sequential recommendation models such as SASRec capture users' evolving preferences from interaction histories~\cite{kang2018self}, while feature-interaction models such as DCNv2 provide efficient mechanisms for learning complex relationships among ranking features~\cite{wang2021dcn}. 
More recently, large language models (LLMs) have emerged as another approach to reranking because they can directly reason over textual representations of user histories and candidate items~\cite{hou2024large,luo2025recranker}.

Despite their flexibility, LLM-based reranking introduces an important tension between \emph{recommendation quality} and \emph{inference efficiency}.
Existing approaches formulate ranking in pointwise, pairwise, or listwise forms~\cite{luo2025recranker,chao2024make}.
A pointwise LLM reranker independently estimates the relevance of each candidate, enabling fine-grained item-level judgments but requiring computation to grow with the number of candidates.
Listwise reranking instead evaluates multiple candidates jointly, reducing the number of model invocations but requiring the model to reason over increasingly long and complex candidate lists.
Prior work has noted both the computational inefficiency of pointwise and pairwise LLM ranking and the challenges faced by listwise approaches in accurately modeling ordering relationships~\cite{chao2024make,qin2024large}.
These tradeoffs are particularly consequential in practical recommendation systems, where rerankers may need to process tens or hundreds of candidates under latency constraints.

Recently, \emph{decision-oriented inference} has emerged as a specialized approach to predictive decision-making, offering an alternative to conventional scoring methods and generative LLM-based approaches.
Representative systems include TypeSafe Jev~\cite{typesafe2026jev} and OpenAI Decisions~\cite{openai_decisions} (hereafter Decisions), which provide structured interfaces for evaluating predefined alternatives and returning associated probabilities.
Unlike conventional generative reranking, which relies on language-model outputs or token probabilities to derive rankings, these systems directly formulate candidate selection as a structured decision problem.
This formulation naturally accommodates recommendation reranking: a user's interaction history provides decision context, retrieved items define a finite set of alternatives, and the resulting probabilities can serve as ranking scores.

However, whether this specialization translates into distinctive practical advantages remains unclear.
Moreover, different decision-oriented implementations may exhibit substantially different characteristics despite supporting similar decision formulations.
This raises a broader question: \emph{does decision-oriented inference provide practical benefits that extend beyond conventional performance--efficiency trade-offs, or does its value primarily depend on the characteristics of individual implementations?}

In this work, we systematically investigate two decision-oriented systems, Jev and Decisions, through personalized recommendation reranking.
Rather than evaluating end-to-end retrieval, we construct hard candidate sets in which the held-out next item is paired with behaviorally plausible negatives retrieved by conventional retriever models, allowing us to evaluate reranking effectiveness independently of retrieval failures.
We conduct experiments across three recommendation domains from Amazon Reviews 2023, namely Movies and TV, Video Games, and Books~\cite{hou2026bridging}, with candidate-set sizes ranging from $K=20$ to $K=200$.
Our baselines include recommendation-specific models, SASRec~\cite{kang2018self} and DCNv2~\cite{wang2021dcn}, as well as pointwise and listwise LLM-based rerankers using Qwen2.5 7B and 32B~\cite{qwen2025qwen25technicalreport} and GPT-based models.

Specifically, our study addresses three research questions.
First, we investigate whether decision-oriented rerankers achieve competitive recommendation effectiveness relative to conventional and LLM-based approaches.
Second, we examine their robustness to candidate presentation order by evaluating identical candidate sets under different permutations, distinguishing aggregate performance variation from changes in individual recommendation outcomes.
Third, we investigate inference efficiency and scalability by analyzing observed latency across candidate-set sizes and comparing the empirical quality--latency trade-offs of different reranking approaches.

Our results show that, in terms of recommendation effectiveness, Jev achieves competitive performance and frequently approaches the strongest LLM-based reranker, while Decisions exhibits a larger effectiveness gap in the primary evaluation.
Both decision-oriented systems generally improve upon their retrieval baselines, although these improvements are not guaranteed across all settings.
Regarding robustness, aggregate NDCG@10 does not differ significantly across candidate permutations, yet individual recommendation outcomes exhibit substantial variation, with Jev generally demonstrating greater stability than Decisions.
Finally, Jev frequently occupies a favorable position on the empirical quality--latency Pareto frontier, offering competitive recommendation quality at considerably lower observed latency than stronger LLM-based rerankers, although such advantages are not uniformly shared by Decisions.

By systematically characterizing its effectiveness, robustness, and efficiency, our study provides empirical evidence for understanding both the potential and the limitations of decision-oriented inference in recommendation reranking.

\section{Related Work}
\label{sec:related_work}

\subsection{Sequential and Text-Aware Recommendation}
Sequential recommendation models user interaction histories to predict future preferences. 
Transformer-based approaches such as SASRec~\cite{kang2018self} and BERT4Rec~\cite{sun2019bert4rec} capture dependencies among historical interactions and have become widely used sequential recommendation architectures.
More recent text-aware methods incorporate semantic item information beyond learned item IDs.
UniSRec~\cite{hou2022towards} learns transferable item and sequence representations from textual descriptions, while RecFormer~\cite{li2023text} represents items and interaction sequences through language representations.
These methods demonstrate the value of combining behavioral and semantic information for recommendation.
Our study focuses specifically on the reranking stage and uses SASRec and DCNv2~\cite{wang2021dcn} as representative recommendation-specific baselines.

\subsection{Large Language Models for Recommendation Reranking}
Large language models have increasingly been applied to recommendation because they can reason directly over textual representations of users and items~\cite{wu2024survey,lin2025can,lyu2024llm}.
Particularly relevant to our setting, \citet{hou2024large} formulate recommendation as ranking a retrieved candidate set conditioned on a user's interaction history, demonstrating the potential of LLMs for zero-shot recommendation ranking.
Subsequent work has explored pointwise, pairwise, and listwise ranking formulations, including instruction-tuned approaches such as RecRanker~\cite{luo2025recranker}.
These formulations exhibit different computational characteristics: pointwise ranking evaluates candidates individually, whereas listwise ranking considers multiple candidates jointly.
Our study compares pointwise and listwise LLM-based rerankers under identical candidate sets and examines how their recommendation quality and observed latency change as candidate-set size increases.

\subsection{Structured Decision Making}
Recent work has explored the use of large pretrained models for structured decision making rather than only language generation.
Decision Transformer~\cite{chen2021decision} formulates reinforcement learning as conditional sequence modeling. 
Gato~\cite{reed2022generalist} similarly applies autoregressive sequence modeling to a multimodal, multitask policy that can emit both text and action tokens, while RT-2~\cite{zitkovich2023rt} co-fine-tunes pretrained vision-language models to produce robotic actions represented as tokens. 
More recently, decision-oriented inference systems, such as TypeSafe Jev~\cite{typesafe2026jev} and OpenAI Decisions~\cite{openai_decisions}, have introduced structured interfaces for predictive decision-making over predefined alternatives.
Our work empirically investigates how this decision-oriented formulation behaves relative to recommendation-specific models and the evaluated LLM rerankers, with particular attention to recommendation effectiveness, candidate-order robustness, and quality--latency trade-offs.

\section{Problem Formulation}\label{sec:problem_formulation}
We study controlled candidate reranking in a two-stage recommendation setting. 
Our objective is not to introduce a new recommendation architecture, but to investigate how decision-oriented, LLM-based, and recommendation-specific models compare when reranking the same behaviorally plausible candidate sets.

\subsection{Two-Stage Recommendation}
\label{sec:two_stage_recommendation}
Let $\mathcal{U}$ denote the set of users and $\mathcal{I}$ the item catalog. For each user $u \in \mathcal{U}$, we observe an ordered interaction history
\begin{equation}
    H_{u}=\left[i_{u,1}, i_{u,2}, \ldots, i_{u, T_{u}}\right]
\end{equation}
where $i_{u,t}\in\mathcal{I}$ is an item previously interacted with by user $u$. The recommendation task is to rank candidate items according to their likelihood of being the user's next interaction.

We adopt a two-stage pipeline. A retrieval model first produces a ranked list of candidate items from the full catalog
\begin{equation}
    R_{u}=\left[r_{u,1}, r_{u,2}, \ldots \right]
\end{equation}
where items are ordered according to the retrieval score. In our experiments, SASRec~\cite{kang2018self} serves as the retrieval model. We additionally use LightGCN~\cite{he2020lightgcn} for the generalizability test.

A second-stage reranker then receives a smaller candidate set
\begin{equation}
    C_{u}^{K}=\{c_{u,1}, c_{u,2}, \ldots, c_{u,K}\}
\end{equation}
and produces a new ranking
\begin{equation}
    \pi_{u}^{(K)}=\text{Rank}\left( C_{u}^{(K)} | H_{u} \right)
\end{equation}
Here, $K$ controls the size of the reranking problem. We study $K \in \{20, 50, 100, 200\}$. 
The ground-truth next item for user $u$ is denoted by $i_{u}^{+}$. Recommendation effectiveness is determined by the position assigned to $i_{u}^{+}$ in $\pi_{u}^{(K)}$.

\subsection{Controlled Hard Candidate Reranking}
\label{sec:controlled_hard_candidate_reranking}
Candidate construction can substantially affect the difficulty of reranking.
Randomly sampled negatives can yield artificially easy ranking problems because many sampled items may be semantically unrelated to the user's interests. We therefore focus on a controlled hard candidate setting in which negative items are drawn from highly ranked retrieval results.

We first sample a fixed set of valid test users and retain those for whom the held-out item appears within SASRec’s top 200 predictions.
The eligible evaluation population is therefore
\begin{equation}
    \mathcal{U}_{\text{eval}}=\left\{ u \in \mathcal{U}: \text{rank}_{R_{u}}\left( i_{u}^{+}\right) \leq K_{\text{max}} \right\}
    \label{eq:eligible}
\end{equation}
where $K_{\text{max}}=200$.
For each eligible user and candidate size $K$, we construct
\begin{equation}
    C_{u}^{K} = \left\{ i_{u}^{+} \right\} \cup N_{u}^{(K-1)}
    \label{eq:candidate}
\end{equation}
where $N_{u}^{(K-1)}$ contains $K-1$ highly ranked non-target items from the retrieval model.
This construction ensures that every evaluated reranker receives exactly one relevant item together with behaviorally plausible competing items. It also separates the reranking problem from retrieval failure: all evaluated methods are compared only when the relevant item has already been successfully retrieved within the top 200 candidates.
It is important to note that this setting should not be interpreted as directly reranking the retriever's top $K$ items. For some users, the ground-truth item may have an original retrieval rank larger than $K$. Our goal is instead to create a controlled candidate set containing the ground-truth item and strong behavioral negatives while keeping the candidate set identical across reranking methods.

\subsection{Reranking Paradigms}
\label{sec:reranking_paradigms}
Given the same interaction history $H_u$ and candidate set $C_u^{(K)}$, different model families can produce ranking scores in different ways. Recommendation-specific models primarily rely on learned item representations
and behavioral interaction patterns.
In our experiments, SASRec and DCNv2 serve as representative
recommendation-specific baselines.

The evaluated language-based rerankers additionally operate on textual
representations of user histories and candidate items.
Let $x_{i}$ denote the textual representation of item $i$, constructed from available metadata such as its title, description, and category information. The semantic user context is represented by the textual descriptions of the user's recent interactions 
\begin{equation}
X_{u} =\left[x_{i_{u,T_{u}-L+1}},\ldots,x_{i_{u,T_{u}}}\right]
\end{equation}
where $L$ denotes the number of historical interactions exposed to the semantic reranker.
A pointwise reranker estimates the relevance of each candidate independently,
\begin{equation}
    s_{u,i}=f\left( X_{u}, x_{i}\right)
\end{equation}

A listwise reranker instead considers the complete candidate set jointly,
\begin{equation}
\mathbf{s}_{u} = f\left( X_{u}, \left\{ x_{i}: i\in C_{u}^{(K)}\right\}\right)
\end{equation}

The final ranking is obtained by sorting candidates according to the resulting relevance scores.

\subsection{Reranking as a Structured Decision Problem}\label{sec:decision_based_reranking}
We additionally formulate candidate reranking as a structured decision problem.
For a user $u$, we define the decision state as the user's recent interaction history $S_{u}=X_{u}$, and treat each candidate item $i\in C_{u}^{(K)}$ as an available decision alternative. A decision-oriented model then estimates a distribution over candidate choices
\begin{equation}
    P\left( i | S_{u}, C_{u}^{(K)}\right), i\in C_{u}^{(K)}
\end{equation}
The resulting ranking is 
\begin{equation}
    \pi_{u}^{(K)}=\text{argsort}^{\downarrow}_{i\in C_{u}^{(K)}}P\left( i | S_{u}, C_{u}^{(K)}\right)
\end{equation}
In our empirical study, we instantiate this formulation using Jev and Decisions. For both systems, the user's recent interaction history provides the decision context, while candidate item descriptions define the available alternatives. The returned candidate-level probabilities are directly used as reranking scores.

\section{Experimental Framework}
\label{sec:exp}

We design a controlled empirical evaluation to characterize how decision-oriented models, recommendation-specific models, Qwen-based and GPT-based LLM rerankers behave under the same candidate reranking setting.

\begin{table}[t]
\centering
\caption{Statistics of the Amazon Reviews 2023 datasets used in our experiments. The table reports the number of users, items, and interactions in each domain before constructing the controlled reranking evaluation set.}
\label{tab:dataset}
\adjustbox{width=\linewidth}{
\begin{tabular}{lrrr}
\toprule
Domain & \# Users & \# Items & \# Interactions \\
\midrule
Movies and TV & 657,203 & 197,943 & 7,441,129 \\
Video Games   & 94,762 & 25,612 & 814,586 \\
Books         & 776,370 & 495,063 & 9,488,297 \\
\bottomrule
\end{tabular}
}
\end{table}

\subsection{Datasets and Preprocessing}
\label{sec:dataset_and_preprocessing}

We conduct experiments on three domains from the Amazon Reviews 2023 benchmark~\cite{hou2026bridging}: Movies and TV, Video Games, and Books.
We use the 5-core leave-last-out configurations, \texttt{5core\_last\_out\_w\_his\_\{domain\}}, which retain users and items with at least five interactions and provide temporally ordered user histories together with held-out next-item interactions.
We use the predefined dataset splits without further resplitting.

For each user, the held-out item in the test split is treated as the ground-truth next interaction $i_u^{+}$.
Recommendation-specific models operate on item identifiers and behavioral histories.
For language-based methods, each item is represented using available textual metadata, including the title, main category, category information, and up to the first 250 characters of the item description.
We expose the 10 most recent historical interactions to all language-based methods and use the same textual representation procedure across these methods.

We use the same preprocessing and evaluation protocol across all domains. Dataset-specific statistics, including the number of users, items, and
interactions, are reported in Table~\ref{tab:dataset}.

\subsection{Candidate Retrieval and Controlled Hard Candidate Construction}\label{sec:candidate_construction}
We adopt SASRec as the first-stage retrieval model.
SASRec is trained independently for each domain using the corresponding training split and produces a relevance score over the item catalog for each test user. Implementation details are in Appendix~\ref{sec:impl_sasrec}.

To separate reranking performance from retrieval failure, we restrict the controlled reranking evaluation to users whose ground-truth next item is retrieved within the top 200 SASRec predictions, as shown in Equation~\ref{eq:eligible}.
After applying this eligibility criterion, we evaluate 954 users for Movies and TV, 1,000 users for Video Games, and 626 users for Books. For Video Games, where more than 1,000 eligible users are available, we cap the evaluation at 1,000 users for computational efficiency. 
For each domain and candidate-set size, all methods are evaluated on the common set of users with successful inference results, ensuring fair comparisons within each setting. Paired statistical significance tests are conducted on the same overlapping users.
We then construct candidate sets with $K\in \{20,50,100,200\}$.
For each eligible user and candidate size $K$, the candidate set consists of the ground-truth next item together with $K-1$ highly ranked non-target items from SASRec, as shown in Equation~\ref{eq:candidate}. 
The negative candidates are therefore behaviorally plausible alternatives rather than randomly sampled items.

Candidate membership is fixed before evaluating any reranker, and all methods operate over exactly the same candidate set for a given user and $K$. Candidate order is randomized deterministically so that the input does not reveal the SASRec ranking.

\subsection{Compared Methods}\label{sec:baselines}

We compare Jev and Decisions with recommendation-specific models, Qwen-based pointwise and listwise rerankers, and GPT listwise reranking. These baselines represent different approaches to reranking, including candidate-level relevance scoring, joint candidate ranking, and structured probabilistic choice.

\paragraph{SASRec} 
SASRec~\cite{kang2018self} serves both as the first-stage retriever and as a conventional behavioral recommendation baseline. For the reranking evaluation, we preserve the original SASRec scores of the items in each controlled candidate set and rank the candidates according to these scores.
This baseline measures how much reranking changes recommendation quality relative to the behavioral model used to construct the hard candidates.

\paragraph{DCNv2}
We include DCNv2~\cite{wang2021dcn} as a neural ranking baseline with explicit feature interaction modeling. We represent the user's historical interactions through pooled item embeddings and combine this representation with the embedding of each candidate item. The resulting features are processed by the cross and deep networks to obtain candidate-level relevance scores. DCNv2 is trained separately for each domain and evaluated on the same frozen candidate sets as all other methods. See Appendix~\ref{sec:impl_dcnv2} for more details.

\paragraph{Qwen Pointwise Reranking}
We evaluate pointwise LLM reranking using Qwen2.5 7B Instruct and 32B Instruct~\cite{qwen2025qwen25technicalreport}. Each candidate is independently evaluated given the same textual user history. For candidate $i$, the model predicts whether the user is likely to interact with the item next. If $z_{0}$ and $z_{1}$ denote the logits corresponding to the negative and positive decisions, respectively, we compute
\begin{equation}
    s_{u,i}=\frac{\exp(z_{1})}{\exp(z_{0})+\exp(z_{1})}
\end{equation}
Candidates are ranked according to $s_{u,i}$.
This formulation obtains a separate candidate-level relevance score for each of the $K$ items. 
The prompt and implementation details are provided in Appendices~\ref{sec:pointwise_prompt} and \ref{sec:impl_qwen}, respectively.

\paragraph{Qwen Listwise Reranking}
We additionally evaluate a listwise formulation using the same Qwen2.5 7B Instruct and 32B Instruct backbones. All $K$ candidates are presented jointly, with each candidate assigned a unique label. We obtain the logits corresponding to the candidate labels and normalize them across the available alternatives to derive the ranking. Unlike pointwise reranking, listwise reranking requires a single joint model evaluation per user, but the input becomes increasingly long and the decision space grows as $K$ increases. 
The prompt and implementation details are provided in Appendices~\ref{sec:listwise_prompt} and \ref{sec:impl_qwen}, respectively.
Neither pointwise nor listwise reranking requires free-form text generation.

\paragraph{GPT Listwise Reranking}
We further evaluate listwise reranking using GPT ({\tt gpt-6-sol} and {\tt gpt-6-luna}). As with Qwen listwise reranking, the model receives the same textual user history together with all $K$ candidate items and is asked to rank candidates according to how likely the user is to interact with each item next. However, the hosted GPT interface does not expose the complete next-token vocabulary logits and provides only a limited number of {\tt top-logprob} alternatives. Consequently, the candidate-label scoring strategy used for Qwen cannot recover scores for all candidates when $K$ exceeds this limit. We therefore ask GPT to directly generate a complete permutation of the $K$ candidate indices, which is used as the final reranked list. This formulation supports all evaluated candidate-set sizes, including $K\in \{20,50,100,200\}$. We also explored pointwise GPT reranking in a pilot study, but found it substantially more expensive in serving latency because it requires a separate API request for each candidate; we therefore evaluate GPT using only the listwise formulation in the full experiments across all three datasets. Prompt and implementation details are provided in Appendices~\ref{sec:gpt_input} and \ref{sec:impl_gpt}.

\paragraph{Decisions}
We additionally evaluate OpenAI Decisions using GPT-6-Luna as a structured probabilistic choice baseline. To ensure comparability with Jev, we provide the same textual user interaction history and candidate item descriptions, with each candidate assigned a unique label. We formulate reranking as a single choice question asking which candidate the user is most likely to interact with next. Unlike GPT listwise reranking, which generates an explicit candidate permutation, Decisions directly returns a probability for each candidate. We rank all $K$ candidates in descending order of their returned probabilities, without requiring free-form text generation or separate candidate-level API requests. This formulation closely matches the structured choice setting of Jev, enabling a comparison between their probability-based ranking capabilities. The input formulation and implementation details are provided in Appendices~\ref{sec:decisions_input} and \ref{sec:impl_decisions}, respectively.

\paragraph{Jev}
We formulate reranking as a structured choice problem in which textual descriptions of the user's recent interactions constitute the decision state and the $K$ candidate items define the available alternatives. Jev returns a probability for each candidate:
\begin{equation}
    s_{u,i}^{\text{Jev}}=P\left( i | S_{u}, C_{u}^{(K)} \right)
\end{equation}
which we directly use as the reranking score.
The input formulation is provided in Appendix~\ref{sec:jev_input}.

\subsection{Evaluation Metrics}\label{sec:metric}
Because each evaluation instance contains a single held-out ground-truth item, we evaluate recommendation quality based on the rank assigned to this item.
Our primary metric is NDCG@10. For a ground-truth item appearing at rank $r_{u}$, NDCG@10 reduces to 
\begin{equation}
    \text{NDCG@10}(u)=
    \begin{cases}
        \dfrac{1}{\log_2(r_u+1)}, & r_{u} \leq 10\\
        0, & r_{u} > 10
    \end{cases}
\end{equation}
NDCG@10 rewards methods that place the ground-truth item near the top of the final recommendation list and is used as our principal measure of reranking effectiveness. 

We additionally report Hit Rate@10,
\begin{equation}
    \text{HR@10}(u)=\mathbb{I}[r_{u} \leq 10]
\end{equation}
which measures whether the target appears anywhere among the top 10 recommendations, and Mean Reciprocal Rank (MRR),
\begin{equation}
    \text{MRR}=\frac{1}{|\mathcal{U}_{\text{eval}}|}\sum_{u}\frac{1}{r_{u}}
\end{equation}
which captures the overall position of the ground-truth item.

\subsection{Latency Measurement}
\label{sec:latency_measurement}
In addition to recommendation quality, we evaluate the serving efficiency of each reranker.

For locally executed methods, latency measures the wall-clock time from transferring preconstructed model inputs to the GPU through producing the final ranked candidate list, including model inference, score extraction, and ranking. Data loading, metadata construction, prompt construction, and item-ID preprocessing are excluded. 

All local latency experiments are conducted on a single NVIDIA A800-SXM4-80GB GPU. The same hardware is used for SASRec, DCNv2, and Qwen-based rerankers to ensure consistent measurement across locally executed methods. 
Jev is evaluated using observed hosted API latency and is therefore \textbf{not} hardware normalized with the locally executed models.

For SASRec, latency includes transferring the preconstructed inputs to the GPU, encoding the user history, scoring the $K$ candidate items, ranking the resulting scores, and transferring the final ranking back to the CPU. For DCNv2, latency similarly includes input transfer, candidate scoring, ranking, and result transfer. Input construction, item-ID conversion, disk I/O, and metric computation are excluded from the timed region.

For pointwise LLM reranking, latency includes tokenization, input transfer, the batched forward passes required to score all $K$ candidates, logit and probability extraction, result transfer, and final candidate sorting. For listwise LLM reranking, latency includes tokenization, input transfer, the joint forward pass over the complete candidate set, candidate-label score extraction, result transfer, and final sorting. Prompt-string construction is excluded from the timed region.

For GPT listwise reranking, we measure the observed wall-clock serving latency of the hosted API. The prompt is constructed before timing begins. The timed region starts immediately before the API request and ends after the generated output has been parsed and converted into the final candidate ranking. The measurement therefore includes network communication, provider-side inference and autoregressive generation, response transfer, local parsing, and final ranking construction. Because GPT directly generates the complete candidate permutation, its output length grows with $K$, in contrast to Qwen listwise reranking, which obtains the ranking from candidate-label logits after a single forward pass. GPT latency should therefore be interpreted as observed end-to-end serving latency under the evaluated deployment rather than intrinsic or hardware-normalized model inference time.

For Decisions, we measure the observed latency of a single structured-choice API request per user. The request contains the user interaction history and all $K$ candidate descriptions, and the API returns candidate-level choice probabilities. Following the measurement protocol used for Jev, the primary latency measure records the elapsed time of the successful API request, including network communication, provider-side processing, and response transfer, while excluding local response parsing and candidate sorting. Automatic SDK retries are disabled, and transient failures are handled through explicit retries with exponential backoff. We separately record total request-processing latency, including failed attempts, retry waiting, and response parsing. Implementation details are provided in Appendix~\ref{sec:impl_decisions}.

Jev is accessed through the TypeSafe hosted API. Implementation details are provided in Appendix~\ref{sec:impl_jev}. We record the elapsed time of the successful request as its primary latency measure. Because hosted inference also depends on network communication, gateway routing, and provider availability, we refer to this quantity as \emph{observed serving latency} rather than intrinsic model inference time. We separately record end-to-end latency including failed requests, retry waiting, and subsequent attempts. Accordingly, latency results should be interpreted under our experimental deployment setting rather than as a hardware-normalized comparison of computational complexity.

\subsection{Experimental Questions}\label{sec:exp_questions}
Our experimental framework is designed to answer three questions:
\begin{itemize}[leftmargin=*]
\item \textbf{RQ1: Recommendation Effectiveness.}
Do decision-oriented rerankers offer competitive recommendation quality compared with conventional and LLM-based methods?
\item \textbf{RQ2: Candidate-Order Robustness.}
How stable are decision-oriented rerankers when the same candidates are presented in different orders?
\item \textbf{RQ3: Efficiency and Scalability.}
What quality–latency trade-offs do decision-oriented rerankers offer, and how do they change with candidate-set size?
\end{itemize}

\section{Results}\label{sec:result}

\begin{table*}[t]
\centering
\caption{Recommendation effectiveness on three Amazon Reviews 2023 domains, measured by NDCG@10 ($\uparrow$) across different candidate-set sizes $K$. (P) and (L) denote pointwise and listwise reranking, respectively. Listwise Qwen uses locally accessible logits, whereas GPT exposes only the top 20 token log probabilities, requiring different prompting and scoring procedures (Appendices~\ref{sec:listwise_prompt} and \ref{sec:gpt_input}).  The best and second-best results in each setting are highlighted in \textbf{bold} and \underline{underlined}, respectively.}
\label{tab:rec_effectiveness_ndcg}
\resizebox{\textwidth}{!}{
\begin{tabular}{lcccccccccccc}
\toprule
& \multicolumn{4}{c}{Movies and TV} 
& \multicolumn{4}{c}{Video Games} 
& \multicolumn{4}{c}{Books} 
 \\
\cmidrule(lr){2-5}
\cmidrule(lr){6-9}
\cmidrule(lr){10-13}
Method 
& $K=20$ & $K=50$ & $K=100$ & $K=200$
& $K=20$ & $K=50$ & $K=100$ & $K=200$
& $K=20$ & $K=50$ & $K=100$ & $K=200$\\
\midrule
SASRec & 0.096 & 0.096 & 0.095 & \underline{0.096} & 0.081 & 0.081 & 0.081 & 0.081 & 0.082 & 0.082 & 0.082 & 0.082\\
DCNv2 & 0.172 & 0.107 & 0.084 & 0.075 & 0.157 & 0.090 & 0.065 & 0.054 & 0.187 & 0.102 & 0.076 & 0.062\\
\midrule
Qwen2.5 7B (P) & 0.289 & 0.151 & 0.093 & 0.059 & 0.321 & 0.175 & 0.119 & 0.078 & 0.347 & 0.189 & 0.115 & 0.080\\
Qwen2.5 7B (L) & 0.247 & 0.118 & 0.058 & 0.034 & 0.263 & 0.106 & 0.059 & 0.025 & 0.293 & 0.138 & 0.062 & 0.030\\
Qwen2.5 32B (P) & 0.267 & 0.117 & 0.055 & 0.037 & 0.283 & 0.117 & 0.064 & 0.040 & 0.295 & 0.136 & 0.072 & 0.041\\
Qwen2.5 32B (L) & 0.254 & 0.124 & 0.062 & 0.033 & 0.297 & 0.122 & 0.062 & 0.039 & 0.351 & 0.163 & 0.070 & 0.034\\
GPT-6-Sol (L) & \textbf{0.307} & \textbf{0.178} & \textbf{0.140} & \textbf{0.099} & \textbf{0.380} & \textbf{0.241} & \textbf{0.175} & \textbf{0.122} & \textbf{0.444} & \textbf{0.310} & \textbf{0.225} & \textbf{0.183}\\
GPT-6-Luna (L) & 0.288 & 0.143 & 0.087 & 0.043 & 0.327 & 0.178 & 0.123 & 0.047 & 0.397 & 0.246 & 0.149 & 0.071\\
\midrule
Decisions & 0.262 & 0.132 & 0.087 & 0.053 & 0.331 & 0.197 & 0.128 & 0.093 & 0.361 & 0.227 & 0.168 & 0.121\\
Jev & \underline{0.293} & \underline{0.162} & \underline{0.117} & 0.077 & \underline{0.360} & \underline{0.228} & \underline{0.153} & \underline{0.107} & \underline{0.406} & \underline{0.263} & \underline{0.191} & \underline{0.152}\\
\bottomrule
\end{tabular}}
\end{table*}

\subsection{Recommendation Effectiveness}\label{sec:result_effectiveness}

\paragraph{Overall effectiveness.} 
Table~\ref{tab:rec_effectiveness_ndcg} presents the NDCG@10 performance of decision-oriented rerankers and conventional baselines across three recommendation domains and four candidate-set sizes. 
GPT-6-Sol achieves the highest NDCG@10 in all 12 settings, demonstrating the effectiveness of strong general-purpose LLMs for recommendation reranking. 
Statistical significance tests further show that GPT-6-Sol significantly outperforms Decisions in all 12 settings and Jev in six settings after Holm--Bonferroni correction (Appendix~\ref{sec:sig_test}).
Among the two decision-oriented approaches, Jev consistently outperforms Decisions and achieves the second-highest NDCG@10 in 11 of the 12 settings. Notably, Jev also outperforms all evaluated Qwen variants and GPT-6-Luna across all settings. Decisions remains competitive, outperforming the evaluated Qwen variants in Video Games and Books, although its performance is less consistent in Movies and TV. 
These results demonstrate that decision-oriented inference can achieve competitive recommendation quality, but does not inherently outperform stronger LLM-based alternatives.
We additionally evaluate MRR and HR@10 in Appendix Tables~\ref{tab:rec_effectiveness_mrr} and \ref{tab:rec_effectiveness_hit}, respectively. Their results generally corroborate the NDCG@10 findings.

\paragraph{Impact of candidate-set size.} 
Increasing the candidate-set size generally reduces reranking effectiveness, reflecting the greater difficulty of identifying the relevant item among additional candidates. For example, as $K$ increases from 20 to 200 in Movies and TV, Jev's NDCG@10 decreases from 0.293 to 0.077, while Decisions decreases from 0.262 to 0.053. Similar trends are observed across the other domains and LLM-based rerankers. Although decision-oriented approaches generally improve upon the underlying SASRec ranking, these improvements are not universal. In Movies and TV, Decisions falls below SASRec at $K=100$ and $K=200$, while Jev also underperforms SASRec at $K=200$. Thus, increasing the candidate-set size can diminish the benefits of decision-oriented reranking.

\paragraph{Generalizability across retrievers.}
To examine whether the effectiveness of decision-oriented reranking depends on the underlying retrieval model, we additionally evaluate Jev and Decisions using candidate sets generated by LightGCN~\cite{he2020lightgcn} (implementation details in Appendix~\ref{sec:impl_lightgcn}), alongside the original SASRec-based setting. As shown in Table~\ref{tab:retriever}, both decision-oriented rerankers consistently improve upon their respective retrieval baselines across all three domains at $K=20$, demonstrating that their effectiveness is not limited to a particular retrieval backbone. 
For example, in Books, Decisions and Jev achieve NDCG@10 scores of 0.487 and 0.382 on LightGCN-generated candidates, respectively, substantially exceeding the LightGCN baseline of 0.068. Interestingly, Decisions outperforms Jev across all three domains under both retrievers, contrasting with the primary evaluation results in Table~\ref{tab:rec_effectiveness_ndcg}. Since this experiment uses a restricted set of users eligible under both retrievers, this difference cannot be attributed solely to the retrieval backbone. Rather, it highlights the sensitivity of relative model performance to evaluation populations and candidate-set characteristics. Overall, these findings demonstrate that decision-oriented reranking can generalize across different retrieval backbones, although the relative advantages of individual implementations are not universal.

\begin{table}[t]
    \centering
    \caption{Generalizability of decision-oriented recommendation reranking across SASRec and LightGCN retrievers, measured by NDCG@10 ($\uparrow$) at candidate-set size $K=20$ on three Amazon Reviews 2023 domains. All methods are evaluated on the same overlapping set of eligible users to ensure a fair cross-retriever comparison. The original retriever rankings are included as baselines. }
    \label{tab:retriever}
    \resizebox{\linewidth}{!}{
    \begin{tabular}{llccc}
    \toprule
    Retriever & Reranker & Movies and TV & Video Games & Books \\
    \midrule
\multirow{3}{*}{SASRec} & SASRec & 0.150 & 0.119 & 0.203\\
& Decisions & 0.307 & 0.389 & 0.366\\
& Jev & 0.288 & 0.355 & 0.338\\
\midrule
\multirow{3}{*}{LightGCN} & LightGCN & 0.089 & 0.114 & 0.068\\
& Decisions & 0.360 & 0.367 & 0.487\\
& Jev & 0.305 & 0.324 & 0.382\\
\bottomrule
    \end{tabular}}
\end{table}

\begin{table*}[t]
\centering
\caption{Candidate-order robustness of Jev and Decisions across different candidate-set sizes on Movies and TV. For each user, we evaluate three permutations of the same candidate set (one original and two shuffled). Mean NDCG@10 is averaged across permutations, while Mean Absolute $\Delta$NDCG@10 and Hit@10 Flip Rate are averaged over all permutation pairs. Informative users are those achieving nonzero NDCG@10 under at least one permutation. The informative-user metrics are computed on this subset, with its size reported as \textit{N} Informative. Lower absolute changes and flip rates, and higher unchanged rates, indicate greater robustness.}
\label{tab:order_robustness}
\resizebox{\textwidth}{!}{
\begin{tabular}{lcccccccc}
\toprule
& \multicolumn{4}{c}{Jev} 
& \multicolumn{4}{c}{Decisions} \\
\cmidrule(lr){2-5}
\cmidrule(lr){6-9}
Metric 
& $K=20$ & $K=50$ & $K=100$ & $K=200$
& $K=20$ & $K=50$ & $K=100$ & $K=200$ \\
\midrule
Mean NDCG@10 & 0.288 & 0.162 & 0.113 & 0.077 & 0.265 & 0.134 & 0.086 & 0.054\\
Mean Absolute $\Delta$NDCG@10 & 0.138 & 0.074 & 0.051 & 0.027 & 0.176 & 0.094 & 0.056 & 0.042\\
Hit@10 Flip Rate & 0.268 & 0.149 & 0.095 & 0.048 & 0.349 & 0.172 & 0.094 & 0.075\\
Informative Mean Absolute $\Delta$NDCG@10 & 0.176 & 0.172 & 0.177 & 0.157 & 0.221 & 0.238 & 0.236 & 0.260\\
Informative-User Unchanged Rate & 0.139 & 0.161 & 0.170 & 0.202 & 0.072 & 0.069 & 0.088 & 0.059\\
N Informative & 746 & 409 & 277 & 163 & 762 & 375 & 227 & 152\\
\bottomrule
\end{tabular}}
\end{table*}

\paragraph{RQ1 takeaway.}
Decision-oriented rerankers achieve competitive recommendation effectiveness and consistently improve upon their retrieval baselines in the cross-retriever evaluation. However, they do not consistently outperform the strongest LLM-based reranker, and their relative effectiveness varies with the evaluation setting.

\subsection{Candidate-Order Robustness}\label{sec:order_robustness}
\paragraph{Overall ordering sensitivity.} We investigate the robustness of decision-oriented rerankers to candidate presentation order by evaluating Jev and Decisions under three permutations of identical candidate sets on Movies and TV. Although statistical tests reveal no significant differences in aggregate NDCG@10 across permutations (Appendix~\ref{sec:sig_test_ordering}), Table~\ref{tab:order_robustness} reveals substantial variation at the individual-user level. At $K=20$, Jev and Decisions exhibit mean absolute NDCG@10 changes of 0.138 and 0.176, respectively, with corresponding Hit@10 flip rates of 26.8\% and 34.9\%. Furthermore, among informative users, only 13.9\% for Jev and 7.2\% for Decisions maintain identical NDCG@10 across all permutations. These findings demonstrate that stable aggregate recommendation performance does not necessarily imply stable individual recommendation outcomes.

To distinguish candidate-order sensitivity from inference stochasticity, we additionally repeat inference using identical candidate presentations. The resulting ranking outcomes remain stable across three repeated runs, suggesting that the observed variation across permutations is attributable to changes in candidate presentation rather than stochastic inference behavior.

\paragraph{Differences between decision-oriented implementations.} Jev  generally demonstrates lower ordering sensitivity than Decisions in the primary robustness evaluation. Across all four candidate-set sizes, Jev exhibits lower mean absolute NDCG@10 changes, both overall and among informative users, as well as higher informative-user unchanged rates. Jev also achieves lower Hit@10 flip rates in three of the four settings, with nearly identical rates for the two models at $K=100$. These observations suggest that ordering sensitivity varies across decision-oriented implementations rather than representing a uniform behavioral property of the inference paradigm.

\paragraph{Impact of candidate-set size.} Interestingly, the overall mean absolute NDCG@10 changes decrease as the candidate-set size increases. For example, Jev's mean absolute change decreases from 0.138 at $K=20$ to 0.027 at $K=200$, while Decisions decreases from 0.176 to 0.042. However, this reduction does not necessarily indicate improved robustness. As $K$ increases, the number of informative users decreases substantially, meaning that more users contribute zero changes because the ground-truth item remains outside the top 10 across all permutations. When restricting the analysis to informative users, Jev's mean absolute changes remain between 0.157 and 0.177, while Decisions exhibits changes between 0.221 and 0.260. Thus, ordering sensitivity remains considerable among users for whom the permutations yield at least one successful recommendation. We additionally investigate whether tied candidate probabilities affect these observations. The main patterns remain largely consistent under tie-aware evaluation, as detailed in Appendix~\ref{sec:tie}.

\begin{figure*}[ht]
    \centering

    \ref*{sharedlegend_quality_latency}


\noindent
\begin{minipage}{0.24\textwidth}
    \centering
    \small\bfseries $K=20$
\end{minipage}
\hfill
\begin{minipage}{0.24\textwidth}
    \centering
    \small\bfseries $K=50$
\end{minipage}
\hfill
\begin{minipage}{0.24\textwidth}
    \centering
    \small\bfseries $K=100$
\end{minipage}
\hfill
\begin{minipage}{0.24\textwidth}
    \centering
    \small\bfseries $K=200$
\end{minipage}


    \begin{subfigure}[t]{0.24\textwidth}
        \centering
        \begin{tikzpicture}[trim axis left, trim axis right]
            \begin{axis}[
                width=3.3cm,
                height=3.3cm,
                scale only axis,
                xlabel={Latency (s)},
                xlabel style={
    yshift=6pt
},
                ylabel={\shortstack{Movies and TV\\NDCG@10}},
                ylabel style={
                at={(axis description cs:-0.13,0.5)},
                anchor=south,
            },
            label style={font=\small},
                xmin=-0.2,
                xmax=2.7,
                ymin=0,
                ymax=0.33,
                xtick={0,1,2},
                ytick={0,0.1,0.2,0.3},
                tick label style={font=\scriptsize},
                grid=major,
                line width=0.9pt,
                legend to name=sharedlegend_quality_latency,
                legend columns=5,
                legend style={
                font=\small,
                    draw=none,
                    /tikz/every even column/.append style={
                        column sep=0.5cm
                    }
                },
            ]


\input{tab_fig/fig/mt_pareto_number_input_20}          

            \end{axis}
        \end{tikzpicture}

        \label{fig:mt_20_ndcg_v_latency}
    \end{subfigure}
    \hfill
    \begin{subfigure}[t]{0.24\textwidth}
        \centering
        \begin{tikzpicture}[trim axis left, trim axis right]
            \begin{axis}[
                width=3.3cm,
                height=3.3cm,
                scale only axis,
                xlabel={Latency (s)},
                xlabel style={
    yshift=6pt
},              
xmin=-0.5,
xmax=6.8,
                ymin=0,
                ymax=0.19,
                xtick={0,2,4,6},
                ytick={0,0.05,0.10,0.15},
               scaled y ticks=false,
                yticklabel style={
                    /pgf/number format/fixed,
                },
                tick label style={font=\scriptsize},
                label style={font=\small},
                grid=major,
                line width=0.9pt,
            ]

\input{tab_fig/fig/mt_pareto_number_input_50} 
   
            \end{axis}
        \end{tikzpicture}

        \label{fig:mt_50_ndcg_v_latency}
    \end{subfigure}
    \hfill
    \begin{subfigure}[t]{0.24\textwidth}
        \centering
        \begin{tikzpicture}[trim axis left, trim axis right]
            \begin{axis}[
                width=3.3cm,
                height=3.3cm,
                scale only axis,
                xlabel={Latency (s)},
                xlabel style={
    yshift=6pt
},
                xmin=-1,
                xmax=14,
                ymin=0,
                ymax=0.15,
                xtick={0,4,8,12},
                ytick={0,0.04,0.08,0.12},
                scaled y ticks=false,
                yticklabel style={
                    /pgf/number format/fixed,
                },
                tick label style={font=\scriptsize},
                label style={font=\small},
                grid=major,
                line width=0.9pt,
            ]
            \input{tab_fig/fig/mt_pareto_number_input_100}

            \end{axis}
        \end{tikzpicture}

        \label{fig:mt_100_ndcg_v_latency}
    \end{subfigure}
    \hfill
    \begin{subfigure}[t]{0.24\textwidth}
        \centering
        \begin{tikzpicture}[trim axis left, trim axis right]
            \begin{axis}[
                width=3.3cm,
                height=3.3cm,
                scale only axis,
                xlabel={Latency (s)},
                xlabel style={
    yshift=6pt
},
                xmin=-3,
                xmax=28,
                ymin=0,
                ymax=0.105,
                xtick={0,10,20},
                ytick={0,0.05,0.10},
                scaled y ticks=false,
                yticklabel style={
                    /pgf/number format/fixed,
                },
                tick label style={font=\scriptsize},
                label style={font=\small},
                grid=major,
                line width=0.9pt,
            ]

\input{tab_fig/fig/mt_pareto_number_input_200} 
            \end{axis}
        \end{tikzpicture}

        \label{fig:mt_200_ndcg_v_latency}
    \end{subfigure}
\hfill
    \begin{subfigure}[t]{0.24\textwidth}
        \centering
        \begin{tikzpicture}[trim axis left, trim axis right]
            \begin{axis}[
                width=3.3cm,
                height=3.3cm,
                scale only axis,
                xlabel={Latency (s)},
                xlabel style={
    yshift=6pt
},
                ylabel={\shortstack{Video Games\\NDCG@10}},
                ylabel style={
                at={(axis description cs:-0.13,0.5)},
                anchor=south,
            },
            label style={font=\small},
                xmin=-0.5,
                xmax=6.5,
                ymin=0,
                ymax=0.42,
                xtick={0,2,4,6},
                ytick={0,0.1,0.2,0.3,0.4},
                tick label style={font=\scriptsize},
                grid=major,
                line width=0.9pt,
            ]


\input{tab_fig/fig/vg_pareto_number_input_20}    

            \end{axis}
        \end{tikzpicture}

        \label{fig:vg_20_ndcg_v_latency}
    \end{subfigure}
    \hfill
    \begin{subfigure}[t]{0.24\textwidth}
        \centering
        \begin{tikzpicture}[trim axis left, trim axis right]
            \begin{axis}[
                width=3.3cm,
                height=3.3cm,
                scale only axis,
                xlabel={Latency (s)},
                xlabel style={
    yshift=6pt
},
                xmin=-1,
                xmax=16.5,
                ymin=0,
                ymax=0.26,
                xtick={0,4,8,12,16},
                ytick={0,0.1,0.2},
               scaled y ticks=false,
                yticklabel style={
                    /pgf/number format/fixed,
                },
                tick label style={font=\scriptsize},
                label style={font=\small},
                grid=major,
                line width=0.9pt,
            ]

\input{tab_fig/fig/vg_pareto_number_input_50}   
            \end{axis}
        \end{tikzpicture}

        \label{fig:vg_50_ndcg_v_latency}
    \end{subfigure}
    \hfill
    \begin{subfigure}[t]{0.24\textwidth}
        \centering
        \begin{tikzpicture}[trim axis left, trim axis right]
            \begin{axis}[
                width=3.3cm,
                height=3.3cm,
                scale only axis,
                xlabel={Latency (s)},
                xlabel style={
    yshift=6pt
},
                xmin=-2,
                xmax=32,
                ymin=0,
                ymax=0.19,
                xtick={0,10,20,30},
                ytick={0,0.05,0.1,0.15},
                scaled y ticks=false,
                yticklabel style={
                    /pgf/number format/fixed,
                },
                tick label style={font=\scriptsize},
                label style={font=\small},
                grid=major,
                line width=0.9pt,
            ]

\input{tab_fig/fig/vg_pareto_number_input_100}   
 
            \end{axis}
        \end{tikzpicture}

        \label{fig:vg_100_ndcg_v_latency}
    \end{subfigure}
    \hfill
    \begin{subfigure}[t]{0.24\textwidth}
        \centering
        \begin{tikzpicture}[trim axis left, trim axis right]
            \begin{axis}[
                width=3.3cm,
                height=3.3cm,
                scale only axis,
                xlabel={Latency (s)},
                xlabel style={
    yshift=6pt
},
                xmin=-5,
                xmax=65,
                ymin=0,
                ymax=0.13,
                xtick={0,20,40,60},
                ytick={0,0.05,0.10},
                scaled y ticks=false,
                yticklabel style={
                    /pgf/number format/fixed,
                },
                tick label style={font=\scriptsize},
                label style={font=\small},
                grid=major,
                line width=0.9pt,  
            ]
\input{tab_fig/fig/vg_pareto_number_input_200}   
         
            \end{axis}
        \end{tikzpicture}

        \label{fig:vg_200_ndcg_v_latency}
    \end{subfigure}
\hfill
    \begin{subfigure}[t]{0.24\textwidth}
        \centering
        \begin{tikzpicture}[trim axis left, trim axis right]
            \begin{axis}[
                width=3.3cm,
                height=3.3cm,
                scale only axis,
                xlabel={Latency (s)},
                xlabel style={
    yshift=6pt
},
                ylabel={\shortstack{Books\\NDCG@10}},
                ylabel style={
                at={(axis description cs:-0.13,0.5)},
                anchor=south,
            },
            label style={font=\small},
                xmin=-0.5,
                xmax=6.7,
                ymin=0,
                ymax=0.5,
                xtick={0,2,4,6},
                ytick={0,0.1,0.2,0.3,0.4,0.5},
                tick label style={font=\scriptsize},
                grid=major,
                line width=0.9pt,
            ]

\input{tab_fig/fig/b_pareto_number_input_20}
         
            \end{axis}
        \end{tikzpicture}

        \label{fig:b_20_ndcg_v_latency}
    \end{subfigure}
    \hfill
    \begin{subfigure}[t]{0.24\textwidth}
        \centering
        \begin{tikzpicture}[trim axis left, trim axis right]
            \begin{axis}[
                width=3.3cm,
                height=3.3cm,
                scale only axis,
                xlabel={Latency (s)},
                xlabel style={
    yshift=6pt
},
                xmin=-1,
                xmax=17,
                ymin=0,
                ymax=0.33,
                xtick={0,5,10,15},
                ytick={0,0.1,0.2,0.3},
               scaled y ticks=false,
                yticklabel style={
                    /pgf/number format/fixed,
                },
                tick label style={font=\scriptsize},
                label style={font=\small},
                grid=major,
                line width=0.9pt,
            ]

\input{tab_fig/fig/b_pareto_number_input_50}
            \end{axis}
        \end{tikzpicture}

        \label{fig:b_50_ndcg_v_latency}
    \end{subfigure}
    \hfill
    \begin{subfigure}[t]{0.24\textwidth}
        \centering
        \begin{tikzpicture}[trim axis left, trim axis right]
            \begin{axis}[
                width=3.3cm,
                height=3.3cm,
                scale only axis,
                xlabel={Latency (s)},
                xlabel style={
    yshift=6pt
},
                xmin=-2,
                xmax=34,
                ymin=0,
                ymax=0.24,
                xtick={0,10,20,30},
                ytick={0,0.1,0.2},
                scaled y ticks=false,
                yticklabel style={
                    /pgf/number format/fixed,
                },
                tick label style={font=\scriptsize},
                label style={font=\small},
                grid=major,
                line width=0.9pt,
            ]
\input{tab_fig/fig/b_pareto_number_input_100}
            \end{axis}
        \end{tikzpicture}

        \label{fig:b_100_ndcg_v_latency}
    \end{subfigure}
    \hfill
    \begin{subfigure}[t]{0.24\textwidth}
        \centering
        \begin{tikzpicture}[trim axis left, trim axis right]
            \begin{axis}[
                width=3.3cm,
                height=3.3cm,
                scale only axis,
                xlabel={Latency (s)},
                xlabel style={
    yshift=6pt
},
                xmin=-3,
                xmax=67,
                ymin=0,
                ymax=0.2,
                xtick={0,20,40,60},
                ytick={0,0.10,0.20},
                scaled y ticks=false,
                yticklabel style={
                    /pgf/number format/fixed,
                },
                tick label style={font=\scriptsize},
                label style={font=\small},
                grid=major,
                line width=0.9pt,
            ]
\input{tab_fig/fig/b_pareto_number_input_200}
            \end{axis}
        \end{tikzpicture}

        \label{fig:b_200_ndcg_v_latency}
    \end{subfigure}
\vspace{-3mm}
\caption{Quality--latency trade-off across domains and candidate-set sizes.
Each panel shows NDCG@10 against median observed per-user inference latency (seconds), with rows representing recommendation domains and columns representing candidate-set sizes $K \in \{20,50,100,200\}$. Black lines connect the empirical Pareto-optimal methods, for which no other evaluated method achieves both lower latency and higher NDCG@10. Locally executed methods are evaluated on a single NVIDIA A800-SXM4-80GB GPU, whereas hosted methods are measured using observed API latency, including network communication and provider-side processing. Consequently, hosted and local latencies are not hardware-normalized. Detailed latency measurement protocols are provided in Appendix~\ref{sec:implementation_details}.
}
    \label{fig:quality_latency}
\end{figure*}

\paragraph{RQ2 takeaway.} 
Although aggregate NDCG@10 does not differ significantly across candidate permutations, individual recommendation outcomes exhibit substantial variation. Jev generally demonstrates greater ordering stability than Decisions, but neither model is invariant to candidate presentation order. These findings remain largely consistent under tie-aware evaluation, suggesting that decision-oriented inference does not inherently guarantee robust ranking behavior.

\subsection{Efficiency and Scalability}\label{sec:latency_and_scalability}

\paragraph{Quality--latency trade-off.} 
Figure~\ref{fig:quality_latency} compares recommendation effectiveness and observed inference latency across three domains and four candidate-set sizes. As expected, conventional recommendation models exhibit substantially lower latency than LLM-based and decision-oriented rerankers, although they generally achieve lower NDCG@10. 
GPT-6-Sol consistently delivers the highest recommendation quality but incurs considerably higher inference latency than Jev and Decisions. Interestingly, Jev frequently lies on the empirical quality--latency Pareto frontier, achieving competitive recommendation effectiveness at substantially lower latency than GPT-6-Sol. This suggests that Jev can provide a favorable compromise between recommendation quality and inference latency. In contrast, Decisions generally exhibits lower recommendation effectiveness than Jev, often accompanied by shorter inference latency, representing a different operating point. Nevertheless, these advantages are not universal. In Movies and TV at $K=200$, for example, SASRec achieves higher NDCG@10 than both decision-oriented models while requiring substantially less inference time.

\paragraph{Scalability with candidate-set size.} The evaluated methods exhibit different latency scaling patterns as the candidate-set size increases. As detailed in Appendix~\ref{appendix_sec:rq3}, the median per-user latency of Jev ranges from approximately 152--155 ms at $K=20$ to 198--293 ms at $K=200$ across the three domains. Decisions exhibits even lower latency, ranging from 92--137 ms at $K=20$ to 161--196 ms at $K=200$. In contrast, GPT-6-Sol's median latency increases from 1.27--1.55 s to 6.38--9.02 s over the same range, while pointwise Qwen rerankers incur substantially greater latency at larger candidate-set sizes. These findings suggest that Jev and Decisions exhibit favorable observed latency scaling, although Decisions achieves this efficiency at the expense of lower recommendation quality. Since hosted API latency includes network communication and provider-side processing, these results characterize practical deployment latency rather than hardware-normalized computational efficiency.

\paragraph{RQ3 takeaway.} 
Jev frequently achieves a favorable quality--latency trade-off, offering competitive recommendation effectiveness at considerably lower observed latency than stronger LLM-based rerankers. However, neither Jev nor Decisions consistently dominates alternative methods across candidate-set sizes and domains.

\section{Discussion}\label{sec:discussion}

Our findings raise an important question about the practical value of decision-oriented inference: does specializing inference for structured decisions translate into distinctive predictive advantages? Across the evaluated recommendation reranking settings, the evidence suggests a nuanced answer. Both Jev and Decisions achieve competitive recommendation effectiveness and generally improve upon conventional retrieval baselines. However, neither consistently surpasses the strongest LLM-based reranker, nor does the decision-oriented formulation guarantee stable ranking outcomes. Their principal advantages appear to depend on the particular implementation and operating conditions rather than arising universally from decision-oriented specialization.

These observations distinguish the value of a specialized decision interface from evidence of inherently superior predictive capability. Expressing a task as a structured choice and obtaining candidate-level probabilities may offer practical advantages for system integration, but such functionality does not necessarily imply better recommendation quality or robustness. Our results therefore challenge the assumption that specialization for decision-making alone establishes an effectiveness advantage.

\subsection{Implications for Recommendation Systems}

From a practical perspective, our results suggest that decision-oriented rerankers should be selected according to application requirements rather than assumed advantages of the inference paradigm. When maximizing recommendation effectiveness is the primary objective, stronger general-purpose LLM rerankers such as GPT-6-Sol may be preferable. When inference latency is an important constraint, Jev can provide a favorable compromise, achieving competitive recommendation quality at substantially lower observed latency. Decisions offers another operating point, generally exhibiting lower latency than Jev but also lower effectiveness in the primary evaluation.

However, the appropriate choice depends on both the recommendation setting and the candidate-set size. In particular, improvements over conventional retrieval baselines are not guaranteed, especially for larger candidate sets. Accordingly, deploying a more sophisticated inference approach is not necessarily beneficial when its additional latency is not accompanied by sufficient improvement in ranking quality.

Our findings also emphasize the importance of evaluating effectiveness, robustness, and efficiency jointly. A model that performs well on aggregate recommendation metrics may still exhibit considerable variation in individual outcomes, while a model with stronger predictive performance may introduce substantial inference overhead. Decision-oriented inference is therefore best understood as an additional option within this multidimensional design space rather than a universally superior replacement for existing approaches.

\subsection{Limitations and Future Directions}

Several limitations qualify our findings. First, our experiments focus on recommendation reranking across three domains from Amazon Reviews 2023. Although the domains provide different item characteristics, they share a common data source and task formulation. Furthermore, our controlled candidate sets include the ground-truth item, allowing us to isolate reranking effectiveness but not to assess end-to-end recommendation performance when the relevant item is absent from the retrieved candidates. Differences in eligible and successfully evaluated user populations may also affect comparisons across experimental settings.
Additionally, differences in training-data coverage and knowledge cutoffs across models may influence their familiarity with evaluated items, potentially affecting relative recommendation effectiveness. Although our evaluation provides user interaction histories and candidate descriptions to reduce reliance on pretrained item knowledge, we cannot fully disentangle the effects of knowledge coverage from those of inference methodology. Future studies could investigate generalizability to newly released items and evolving user preferences.

Second, our analysis covers two decision-oriented implementations whose underlying architectures and training procedures are not fully observable. Consequently, the observed performance differences cannot be attributed exclusively to decision-oriented specialization, independently of model capacity, training data, or inference design. Similarly, our candidate-order experiments examine changes in candidate presentation that also modify position-dependent candidate labels. Although repeated identical-input inference produces stable outcomes, the experiments do not isolate positional effects from changes in label assignment.

Third, the efficiency comparisons reflect observed deployment latency rather than hardware-normalized computational efficiency. Hosted API measurements include network communication and provider-side processing, whereas locally executed methods are evaluated on dedicated GPU hardware. The resulting quality--latency comparisons are informative for the evaluated deployment settings but should not be interpreted as intrinsic computational advantages.

Finally, recommendation reranking captures only one aspect of decision-oriented inference. Other applications may involve explicit constraints, multiple objectives, uncertainty-sensitive decisions, or structured actions that are not represented by the present evaluation. Future research could investigate whether decision-oriented specialization provides distinctive benefits in these settings, ideally through more controlled comparisons that help separate the effects of model architecture, training, and inference formulation.

\section{Conclusion}\label{sec:conclusion}

This paper presents a systematic empirical study of decision-oriented inference for recommendation reranking, examining two representative implementations, Jev and Decisions, in terms of recommendation effectiveness, candidate-order robustness, and efficiency and scalability. Through controlled experiments across three recommendation domains and multiple candidate-set sizes, our findings suggest that specialization for decision-making does not automatically translate into superior decision quality or robustness. Instead, the practical value of decision-oriented inference appears to depend on implementation-specific characteristics and application requirements. By identifying both its competitive operating points and its limitations, our study provides empirical evidence for assessing decision-oriented inference as an emerging alternative within the broader landscape of predictive and generative approaches.

\bibliography{custom}

\appendix

\section{Prompt and Input Templates}
\label{sec:prompt_templates}

Across the language-based methods, the user history is represented using the same textual descriptions of the most recent interactions. Candidate item descriptions are constructed using the same metadata preprocessing procedure described in Section~\ref{sec:dataset_and_preprocessing}. Candidate order is deterministically randomized before inference and is held fixed across methods for each user and candidate-set size.

\subsection{Pointwise Qwen Prompt}
\label{sec:pointwise_prompt}

For pointwise reranking, each candidate item is evaluated independently against the same user interaction history. We use the following system and user messages:

\begin{Verbatim}[
    breaklines=true,
    breakanywhere=true,
    fontsize=\small
]
System:
You are a recommendation model. Predict whether the candidate item is likely to be the user's next interaction.

User:
User interaction history:
{history_text}

Candidate item:
{candidate_text}

Is this candidate likely to be the user's next interaction?

Answer with exactly one digit:
1 = likely
0 = unlikely
\end{Verbatim}

Here, \texttt{\{history\_text\}} contains the textual representations of the user's recent interactions, while \texttt{\{candidate\_text\}} contains the textual representation of the candidate item. Rather than relying on generated text, we use the model logits associated with the tokens corresponding to \texttt{0} and \texttt{1}.

\subsection{Listwise Qwen Prompt}
\label{sec:listwise_prompt}

For listwise reranking, all candidates in a candidate set are presented jointly. Each candidate is assigned a unique label, and the model is asked to identify the candidate most likely to correspond to the user's next interaction. We use the following prompt:

\begin{Verbatim}[
    breaklines=true,
    breakanywhere=true,
    fontsize=\small
]
System:
You are a recommendation model. Given a user's interaction history and a set of candidate items, determine which candidate the user is most likely to interact with next.

User:
User interaction history:
{history_text}

Candidate items:
{candidate_text}

Which candidate is the user most likely to interact with next?

Answer with exactly one candidate label.
\end{Verbatim}

Here, \texttt{\{history\_text\}} is constructed in the same way as in the pointwise setting. The \texttt{\{candidate\_text\}} field contains all $K$ candidate items, each associated with a unique candidate label. We obtain the logits corresponding to the candidate labels and normalize them across the available alternatives to derive candidate-level ranking scores. Thus, the model does not need to generate a free-form ranking; instead, ranking is derived directly from the scores assigned to the candidate labels.

\subsection{GPT-based Reranking}
\label{sec:gpt_input}
For GPT-based reranking, we formulate recommendation as next-interaction prediction. Each input contains a user's interaction history and a fixed candidate set. Historical interactions are presented in chronological order from oldest to newest, and each candidate is associated with an integer index. The item representations are constructed using the same metadata representation as in the other LLM-based reranking experiments.

Our GPT and Qwen prompts share the same recommendation objective and input information but differ in their required output format because of differences in model access. For Qwen, which is executed locally, we have direct access to the complete next-token vocabulary logits. We therefore associate each candidate with a single-token label and obtain a listwise ranking by extracting and normalizing the logits of all candidate labels from a single forward pass. This approach naturally supports candidate sets larger than 20.

In contrast, the hosted GPT interface does not expose the complete next-token logit vector and returns only a limited number of {\tt top-logprob} alternatives. Consequently, label-logit ranking cannot recover scores for every candidate when the candidate-set size exceeds this limit (\textit{e.g.,} $K=50,100,200$). We therefore use direct ranking generation for GPT: the model is asked to generate a complete permutation of all candidate indices. Apart from this output mechanism, we keep the task definition, user history, candidate information, and next-interaction objective aligned with the Qwen setting.

For the GPT listwise setting, we use the following prompt template:

\begin{Verbatim}[
    breaklines=true,
    breakanywhere=true,
    fontsize=\small
]
You are a recommendation model. Given a user's interaction history and a set of candidate items, rank the candidates according to how likely the user is to interact with each item next.

User interaction history:
{history_text}

Candidate items:
{candidate_text}

Rank all candidate items from most likely to least likely to be the user's next interaction.

Return only valid JSON with this exact shape:
{"ranking": [candidate_index_1, candidate_index_2, ...]}

Use every candidate index exactly once. Do not output any explanation.
\end{Verbatim}

The generated permutation is directly mapped back to the corresponding item IDs to obtain the final reranked list.

We additionally considered a pointwise GPT formulation in a pilot study. Each candidate was evaluated independently by asking whether it was likely to be the user's next interaction, with the model constrained to output 1 (likely) or 0 (unlikely). Although this formulation permits probability-based candidate scoring from output-token log probabilities, it requires one API request per candidate. We observed substantially higher latency than with the single-call listwise formulation, particularly as $K$ increased. We therefore use GPT listwise reranking for the main experiments on all three datasets.

\subsection{Decisions}
\label{sec:decisions_input}
We formulate recommendation reranking as a structured choice task using the OpenAI Decisions API. To ensure comparability with Jev, we provide the same user interaction history and candidate item descriptions. Specifically, the interaction history is serialized as a JSON object, while each candidate is represented as a choice with a unique identifier (\textit{e.g.,} {\tt C000}, {\tt C001}) and its corresponding textual description. We use a single choice question with the instruction:
\begin{Verbatim}[
    breaklines=true,
    breakanywhere=true,
    fontsize=\small
]
Based on the user's interaction history, which candidate item is the user most likely to interact with next?
\end{Verbatim}
The API returns a probability for each candidate, which we use to construct the final ranking in descending order.

\subsection{Jev Input Formulation}
\label{sec:jev_input}

Jev uses a structured decision interface rather than a conventional system--user prompt format. For each user, the recent interaction history is supplied as the decision state, and candidate items are represented as the available alternatives in a \texttt{choice}-type question. The instruction is as follows:

\begin{Verbatim}[
    breaklines=true,
    breakanywhere=true,
    fontsize=\small
]
Based on the user's interaction history, which candidate item is the user most likely to interact with next?
\end{Verbatim}

Jev returns a probability for each candidate alternative under the \texttt{next\_item} question. We directly use these probabilities as candidate ranking scores.

\section{Implementation Details}
\label{sec:implementation_details}

This section provides additional implementation details for the evaluated methods. Unless otherwise stated, models are trained separately for each recommendation domain. All locally executed inference experiments are conducted on a single NVIDIA A800-SXM4-80GB GPU. Latency is measured according to the protocol described in Section~\ref{sec:latency_measurement}. 

\subsection{SASRec}
\label{sec:impl_sasrec}

We implement SASRec~\cite{kang2018self} as both the first-stage retrieval model and a recommendation-specific baseline. Item identifiers are mapped to learned embeddings, and each user's historical interactions are truncated to a maximum sequence length of 50.

The SASRec model uses an embedding dimension of 64, two self-attention blocks, two attention heads, and a dropout rate of 0.2. Training uses a batch size of 256 with 64 sampled negative items per positive interaction. The model is trained separately for each domain using the corresponding training split. The resulting checkpoint is used both to construct the controlled hard candidate sets and to score candidates in the SASRec baseline.

During inference, the preconstructed user sequence and candidate tensors are transferred to the GPU, after which SASRec encodes the user history, scores all $K$ candidates, and ranks them according to their predicted relevance. The final ranking is transferred back to the CPU before timing ends. Input construction, item-ID conversion, disk I/O, and metric computation are excluded from the timed region.

\subsection{DCNv2}
\label{sec:impl_dcnv2}

We implement DCNv2~\cite{wang2021dcn} as an ID-based neural ranking baseline. Each item is represented using a 64-dimensional learned embedding. The user's historical representation is obtained by pooling the embeddings of the non-padding items in the interaction history and is combined with each candidate-item embedding for ranking.

The ranking network contains a three-layer DCNv2 cross network together with a parallel deep network with hidden dimensions 256 and 128. We use a dropout rate of 0.2 in the deep component. The outputs of the cross and deep components are combined to produce a scalar relevance score for each candidate.

To maintain consistency with SASRec, DCNv2 uses the same domain-specific training split, item vocabulary construction, and negative-sampling procedure. Each positive training instance is paired with 64 sampled negatives. DCNv2 is trained separately for each domain and evaluated on the same frozen candidate sets as all other methods.
DCNv2 uses the same 10 most recent user interactions exposed to the language-based rerankers for reranking evaluation.

During inference, the preconstructed history and candidate tensors are transferred to the GPU, all $K$ candidates are scored in parallel, and the resulting scores are sorted to produce the final ranking. The ranking is transferred back to the CPU before timing ends. Input construction, item-ID conversion, disk I/O, and metric computation are excluded from the timed region.

Both SASRec and DCNv2 are optimized using AdamW with a learning rate of $10^{-3}$ and weight decay of $10^{-5}$, for a maximum of 20 epochs. We select the SASRec checkpoint with the highest validation NDCG@10 and the DCNv2 checkpoint with the lowest validation loss. For both models, training stops early if the corresponding validation criterion does not improve for three consecutive epochs.

\subsection{Qwen Rerankers}
\label{sec:impl_qwen}

We evaluate Qwen2.5 7B Instruct and Qwen2.5 32B Instruct~\cite{qwen2025qwen25technicalreport} in both pointwise and listwise reranking configurations. Both models are executed locally on a single NVIDIA A800-SXM4-80GB GPU using bfloat16 precision. The same textual representation of the user's 10 most recent interactions and the same candidate metadata are used.

For pointwise reranking, each candidate is paired with the user history using the prompt in Appendix~\ref{sec:pointwise_prompt}. Rather than generating a free-form response, we extract the next-token logits associated with the \texttt{0} and \texttt{1} decision tokens and normalize them using a two-way softmax. The probability assigned to \texttt{1} is used as the candidate relevance score. 
Candidate prompts are processed with a batch size of 16, and the final ranking is obtained by sorting candidates according to these scores. In preliminary experiments on a smaller evaluation subset, we varied the pointwise batch size and observed negligible differences in both ranking effectiveness and per-user latency; we therefore use a batch size of 16 throughout the main experiments.

For listwise reranking, all $K$ candidates are presented jointly using the prompt in Appendix~\ref{sec:listwise_prompt}. Each candidate is assigned a unique label corresponding to a single tokenizer token. We extract the next-token logits associated with these candidate labels and normalize them across the candidate set to obtain ranking scores. The complete candidate set is processed in a single model forward pass.

For both formulations, latency includes tokenization, transfer of tokenized inputs to the GPU, model inference, logit and probability extraction, transfer of the resulting scores to the CPU, and final candidate sorting. Prompt-string construction is excluded from the timed region. No sampling-based decoding parameters such as temperature or top-$p$ are used because ranking scores are obtained directly from model logits.

Before running the final experiments, we verify that every pointwise and listwise input remains within the corresponding model context window, including all cases with $K=200$; no evaluated prompt requires truncation.

\subsection{GPT}
\label{sec:impl_gpt}
We evaluate GPT through the OpenAI Responses API using {\tt gpt-6-sol} and {\tt gpt-6-luna}. We disable additional reasoning by setting the reasoning effort to none and use a temperature of 0 for deterministic inference.

For each user, GPT listwise reranking requires a single API request containing the complete interaction history and candidate set. Unlike the Qwen listwise implementation, which obtains candidate scores directly from next-token logits, GPT autoregressively generates the complete candidate ordering. The output is constrained to a JSON object containing a permutation of all candidate indices. We parse the returned indices and map them to their corresponding item IDs. The output-token budget is scaled with the candidate-set size so that the model can return a complete permutation for $K\in \{20,50,100,200\}$.

We use wall-clock latency to characterize the end-to-end serving cost of GPT reranking. Prompt-string construction is performed before timing begins. The timed region starts immediately before the API request and ends after the returned output has been parsed and converted into the final candidate ranking. Thus, the reported latency includes API communication, provider-side inference and autoregressive generation, and local output parsing and ranking construction, while excluding prompt construction and experiment bookkeeping. Because GPT is accessed through a hosted API, this quantity should be interpreted as observed end-to-end serving latency rather than hardware-normalized model inference time.

\subsection{Decisions}
\label{sec:impl_decisions}
We use GPT-6 Luna ({\tt gpt-6-luna}) through the OpenAI Decisions API. For each user, we submit a single request containing the interaction history and a structured choice question over the fixed candidate set. Candidate probabilities are extracted directly from the API response without additional normalization for ranking. We disable automatic SDK retries and implement explicit retry handling for transient API failures, allowing up to six attempts with exponential backoff. The request timeout is set to 120 seconds. For latency evaluation, we measure the elapsed time of the successful API request, excluding response parsing, ranking construction, and retry waiting time, consistent with the latency measurement protocol used for Jev. Failed requests are excluded from the successful evaluation results and recorded separately.

\subsection{Jev}
\label{sec:impl_jev}

Jev is accessed through TypeSafe AI's hosted API using \texttt{jev-1.13.0}. For each user, the textual representation of the 10 most recent interactions is supplied as the decision state, while the $K$ candidate items are represented as alternatives in a \texttt{choice}-type question. The exact request structure is provided in Appendix~\ref{sec:jev_input}.

Jev returns a probability for each candidate alternative, which is used directly as its reranking score without additional calibration or post-processing. Candidate order is deterministically randomized before submission and is identical to that used for the corresponding evaluation cases.

Because Jev is accessed through a hosted service, its underlying hardware, numerical precision, batching strategy, and serving configuration are not available to us. We therefore report observed serving latency rather than intrinsic model inference time, following the protocol described in Section~\ref{sec:latency_measurement}.

\subsection{LightGCN}
\label{sec:impl_lightgcn}
We implement LightGCN~\cite{he2020lightgcn} as an alternative retrieval backbone to examine the generalizability of decision-oriented reranking beyond SASRec. For each recommendation domain, we construct a user--item bipartite graph using unique user--item interactions from the predefined training split, excluding validation and test interactions. Following the standard LightGCN formulation, we employ symmetric degree-normalized graph propagation without nonlinear feature transformations and average node embeddings from all propagation layers, including the initial embeddings. We use 64-dimensional embeddings and three graph propagation layers. Recommendation scores are computed as the inner products of the resulting user and item embeddings.

LightGCN is trained using Bayesian Personalized Ranking (BPR) loss with uniformly sampled negative items that are not connected to the corresponding users in the training graph. We employ the Adam optimizer with a learning rate of $10^{-3}$, an L2 regularization coefficient of $10^{-4}$, and a batch size of 4,096 positive interactions. 
We adopt domain-specific training configurations. For Movies and TV, each epoch consists of 300 optimization steps, with training conducted for at most 200 epochs and early stopping patience of 10 epochs. For Video Games and Books, we use 100 optimization steps per epoch, a maximum of 100 epochs, and early stopping patience of 3 epochs. Each optimization step involves full-graph propagation. Across all domains, the best checkpoint is selected based on validation NDCG@10.

For candidate generation, we score the training item catalog using the learned LightGCN embeddings while masking items appearing in each user's interaction history. Unlike SASRec, which encodes a truncated interaction sequence, LightGCN uses user representations learned from the training interaction graph. Following the SASRec-based evaluation protocol, we consider users whose ground-truth next item appears among the top 200 retrieved items. For each eligible user and candidate-set size $K$, we construct a controlled reranking set comprising the ground-truth item and the $K-1$ highest-scoring negative candidates, followed by deterministic shuffling. We evaluate $K \in \{20,50,100,200\}$ and provide the same 10 most recent historical interactions to the downstream decision-oriented rerankers. To ensure comparability, cross-retriever experiments are conducted on the overlapping set of eligible users, and each retriever's baseline ranking is evaluated on its corresponding fixed candidate set.

\section{Additional Results}
\label{sec:additional_results}

\subsection{RQ1: Recommendation Effectiveness}\label{appendix_sec:rq1}

\subsubsection{MRR and Hit Rate@10}\label{sec:additional_metrics}
To assess whether the observed trends depend on the choice of ranking metric, we additionally evaluate all methods using MRR and Hit Rate@10. The corresponding results are reported in Tables~\ref{tab:rec_effectiveness_mrr} and \ref{tab:rec_effectiveness_hit}.
Overall, the results are consistent with those based on NDCG@10 in the main text. 

\begin{table*}[t]
\centering
\caption{Recommendation effectiveness on three Amazon Reviews 2023 domains, measured by MRR ($\uparrow$) across different candidate-set sizes $K$. (P) and (L) denote pointwise and listwise reranking, respectively. Listwise Qwen uses locally accessible logits, whereas GPT exposes only the top 20 token log probabilities, requiring different prompting and scoring procedures (Appendices~\ref{sec:listwise_prompt} and \ref{sec:gpt_input}).  The best and second-best results in each setting are highlighted in \textbf{bold} and \underline{underlined}, respectively.}
\label{tab:rec_effectiveness_mrr}
\resizebox{\textwidth}{!}{
\begin{tabular}{lcccccccccccc}
\toprule
& \multicolumn{4}{c}{Movies and TV} 
& \multicolumn{4}{c}{Video Games} 
& \multicolumn{4}{c}{Books} 
 \\
\cmidrule(lr){2-5}
\cmidrule(lr){6-9}
\cmidrule(lr){10-13}
Method 
& $K=20$ & $K=50$ & $K=100$ & $K=200$
& $K=20$ & $K=50$ & $K=100$ & $K=200$
& $K=20$ & $K=50$ & $K=100$ & $K=200$\\
\midrule
SASRec & 0.117 & 0.098 & 0.093 & \underline{0.093} & 0.102 & 0.082 & 0.078 & 0.077 & 0.101 & 0.080 & 0.075 & 0.074\\
DCNv2 & 0.152 & 0.104 & 0.085 & 0.074 & 0.143 & 0.091 & 0.069 & 0.056 & 0.159 & 0.099 & 0.074 & 0.061\\
\midrule
Qwen2.5 7B (P) & 0.232 & 0.136 & 0.089 & 0.061 & 0.259 & 0.150 & 0.104 & 0.071 & 0.282 & 0.164 & 0.100 & 0.074\\
Qwen2.5 7B (L) & 0.199 & 0.107 & 0.061 & 0.040 & 0.213 & 0.102 & 0.063 & 0.033 & 0.245 & 0.134 & 0.063 & 0.035\\
Qwen2.5 32B (P) & 0.208 & 0.107 & 0.059 & 0.044 & 0.212 & 0.106 & 0.067 & 0.043 & 0.225 & 0.125 & 0.068 & 0.044\\
Qwen2.5 32B (L) & 0.205 & 0.118 & 0.067 & 0.039 & 0.240 & 0.117 & 0.067 & 0.042 & 0.293 & 0.151 & 0.071 & 0.042\\
GPT-6-Sol (L) & \textbf{0.258} & \textbf{0.168} & \textbf{0.134} & \textbf{0.096} & \textbf{0.315} & \textbf{0.208} & \textbf{0.155} & \textbf{0.113} & \textbf{0.375} & \textbf{0.277} & \textbf{0.210} & \textbf{0.169}\\
GPT-6-Luna (L) & 0.230 & 0.135 & 0.088 & 0.048 & 0.266 & 0.158 & 0.113 & 0.050 & 0.334 & 0.215 & 0.140 & 0.071\\
\midrule
Decisions & 0.215 & 0.127 & 0.086 & 0.053 & 0.269 & 0.174 & 0.113 & 0.084 & 0.299 & 0.198 & 0.155 & 0.113\\
Jev & \underline{0.233} & \underline{0.147} & \underline{0.108} & 0.073 & \underline{0.292} & \underline{0.197} & \underline{0.131} & \underline{0.096} & \underline{0.340} & \underline{0.236} & \underline{0.176} & \underline{0.137}\\
\bottomrule
\end{tabular}}
\end{table*}

\begin{table*}[t]
\centering
\caption{Recommendation effectiveness on three Amazon Reviews 2023 domains, measured by HR@10 ($\uparrow$) across different candidate-set sizes $K$. (P) and (L) denote pointwise and listwise reranking, respectively. Listwise Qwen uses locally accessible logits, whereas GPT exposes only the top 20 token log probabilities, requiring different prompting and scoring procedures (Appendices~\ref{sec:listwise_prompt} and \ref{sec:gpt_input}).  The best and second-best results in each setting are highlighted in \textbf{bold} and \underline{underlined}, respectively.}
\label{tab:rec_effectiveness_hit}
\resizebox{\textwidth}{!}{
\begin{tabular}{lcccccccccccc}
\toprule
& \multicolumn{4}{c}{Movies and TV} 
& \multicolumn{4}{c}{Video Games} 
& \multicolumn{4}{c}{Books} 
 \\
\cmidrule(lr){2-5}
\cmidrule(lr){6-9}
\cmidrule(lr){10-13}
Method 
& $K=20$ & $K=50$ & $K=100$ & $K=200$
& $K=20$ & $K=50$ & $K=100$ & $K=200$
& $K=20$ & $K=50$ & $K=100$ & $K=200$\\
\midrule
SASRec & 0.170 & 0.169 & 0.168 & \textbf{0.169} & 0.161 & 0.161 & 0.161 & 0.161 & 0.166 & 0.166 & 0.165 & 0.166\\
DCNv2 & 0.373 & 0.217 & 0.160 & 0.143 & 0.346 & 0.189 & 0.129 & 0.108 & 0.411 & 0.216 & 0.159 & 0.126\\
\midrule
Qwen2.5 7B (P) & 0.575 & \underline{0.300} & 0.183 & 0.109 & 0.612 & 0.346 & 0.244 & 0.164 & 0.644 & 0.361 & 0.240 & 0.161\\
Qwen2.5 7B (L) & 0.519 & 0.250 & 0.122 & 0.065 & 0.535 & 0.220 & 0.121 & 0.051 & 0.551 & 0.251 & 0.130 & 0.062\\
Qwen2.5 32B (P) & 0.566 & 0.255 & 0.121 & 0.072 & 0.609 & 0.265 & 0.141 & 0.093 & 0.617 & 0.278 & 0.170 & 0.094\\
Qwen2.5 32B (L) & 0.525 & 0.244 & 0.121 & 0.066 & 0.583 & 0.237 & 0.117 & 0.074 & 0.628 & 0.294 & 0.136 & 0.059\\
GPT-6-Sol (L) & 0.566 & \textbf{0.305} & \textbf{0.226} & \underline{0.156} & \textbf{0.667} & \textbf{0.431} & \textbf{0.312} & \textbf{0.216} & \textbf{0.730} & \textbf{0.495} & \textbf{0.348} & \textbf{0.291}\\
GPT-6-Luna (L) & \underline{0.577} & 0.268 & 0.157 & 0.076 & 0.613 & 0.336 & 0.233 & 0.090 & 0.681 & \underline{0.430} & 0.255 & 0.126\\
\midrule
Decisions & 0.524 & 0.248 & 0.158 & 0.096 & 0.618 & 0.360 & 0.244 & 0.175 & 0.645 & 0.406 & 0.274 & 0.198\\
Jev & \textbf{0.587} & 0.299 & \underline{0.211} & 0.131 & \underline{0.659} & \underline{0.415} & \underline{0.298} & \underline{0.197} & \underline{0.693} & \underline{0.430} & \underline{0.300} & \underline{0.241}\\
\bottomrule
\end{tabular}}
\end{table*}

\subsection{Statistical Significance of Recommendation Effectiveness Differences}\label{sec:sig_test}
Table~\ref{tab:sig_test} provides statistical comparisons of NDCG@10 between the two decision-oriented models and GPT-6-Sol. For each dataset and candidate-set size, we calculate the paired per-user NDCG@10 difference, defined as the score of the decision-oriented model minus that of GPT-6-Sol. Statistical significance is assessed using two-sided paired permutation tests, with Holm–Bonferroni correction applied across all 24 comparisons. We additionally report 95\% paired bootstrap confidence intervals to characterize uncertainty in the estimated performance differences.

All observed mean differences are negative, indicating that GPT-6-Sol achieves higher average NDCG@10 than both decision-oriented models in every setting. However, the magnitude and statistical significance of these differences vary. For Jev, six of the 12 comparisons remain statistically significant after correction: Movies and TV at $K=100$ and $K=200$, and Books at all four candidate-set sizes. None of the four comparisons in Video Games reaches the corrected significance threshold. By contrast, Decisions performs significantly worse than GPT-6-Sol in all 12 settings. 

The complementary effectiveness metrics and statistical tests confirm that Jev is a competitive alternative to strong LLM-based reranking, whereas Decisions exhibits a more pronounced effectiveness gap relative to GPT-6-Sol. 

\begin{table*}[ht]
    \centering
    \caption{Statistical significance of NDCG@10 differences between decision-oriented models (Jev and Decisions) and GPT-6-Sol (Listwise) across three datasets and four candidate-set sizes $K$. $\Delta$NDCG@10 denotes the mean paired difference (decision-oriented model minus GPT-6-Sol). We report unadjusted 95\% paired bootstrap confidence intervals (CIs) and p-values from two-sided paired permutation tests with Holm–Bonferroni correction across 24 comparisons. Statistically significant differences ($p < 0.05$) are highlighted in bold.}
    \label{tab:sig_test}
    \resizebox{\textwidth}{!}{
    \begin{tabular}{lccccc}
\toprule
Dataset & $K$ & Comparison & $\Delta$NDCG@10 & 95\% CI & Adjusted p-value\\
\midrule
Movies and TV & 20 & Jev vs GPT-6-Sol & -0.014 & [-0.033, 0.004] & 0.2504\\
Movies and TV & 20 & Decisions vs GPT-6-Sol & \textbf{-0.046} & [-0.065, -0.025] & 0.0024\\
Movies and TV & 50 & Jev vs GPT-6-Sol & -0.016 & [-0.031, -0.002] & 0.1156\\
Movies and TV & 50 & Decisions vs GPT-6-Sol & \textbf{-0.046} & [-0.062, -0.030] & 0.0024\\
Movies and TV & 100 & Jev vs GPT-6-Sol & \textbf{-0.023} & [-0.035, -0.010] & 0.0070\\
Movies and TV & 100 & Decisions vs GPT-6-Sol & \textbf{-0.053} & [-0.068, -0.038] & 0.0024\\
Movies and TV & 200 & Jev vs GPT-6-Sol & \textbf{-0.022} & [-0.033, -0.012] & 0.0024\\
Movies and TV & 200 & Decisions vs GPT-6-Sol & \textbf{-0.046} & [-0.060, -0.033] & 0.0024\\
Video Games & 20 & Jev vs GPT-6-Sol & -0.019 & [-0.036, -0.003] & 0.1075\\
Video Games & 20 & Decisions vs GPT-6-Sol & \textbf{-0.049} & [-0.068, -0.030] & 0.0024\\
Video Games & 50 & Jev vs GPT-6-Sol & -0.013 & [-0.030, 0.004] & 0.2504\\
Video Games & 50 & Decisions vs GPT-6-Sol & \textbf{-0.044} & [-0.063, -0.025] & 0.0024\\
Video Games & 100 & Jev vs GPT-6-Sol & -0.021 & [-0.038, -0.005] & 0.0588\\
Video Games & 100 & Decisions vs GPT-6-Sol & \textbf{-0.047} & [-0.066, -0.028] & 0.0024\\
Video Games & 200 & Jev vs GPT-6-Sol & -0.015 & [-0.030, -0.000] & 0.1305\\
Video Games & 200 & Decisions vs GPT-6-Sol & \textbf{-0.029} & [-0.046, -0.012] & 0.0054\\
Books & 20 & Jev vs GPT-6-Sol & \textbf{-0.038} & [-0.059, -0.017] & 0.0033\\
Books & 20 & Decisions vs GPT-6-Sol & \textbf{-0.083} & [-0.106, -0.059] & 0.0024\\
Books & 50 & Jev vs GPT-6-Sol & \textbf{-0.047} & [-0.068, -0.026] & 0.0024\\
Books & 50 & Decisions vs GPT-6-Sol & \textbf{-0.083} & [-0.106, -0.060] & 0.0024\\
Books & 100 & Jev vs GPT-6-Sol & \textbf{-0.034} & [-0.053, -0.014] & 0.0064\\
Books & 100 & Decisions vs GPT-6-Sol & \textbf{-0.057} & [-0.079, -0.035] & 0.0024\\
Books & 200 & Jev vs GPT-6-Sol & \textbf{-0.032} & [-0.048, -0.015] & 0.0033\\
Books & 200 & Decisions vs GPT-6-Sol & \textbf{-0.062} & [-0.084, -0.041] & 0.0024\\
\bottomrule
    \end{tabular}}
\end{table*}

\subsection{RQ2: Candidate-Order Robustness}\label{appendix_sec:rq2}

\subsubsection{Statistical Significance of Aggregate Performance Differences}\label{sec:sig_test_ordering}

To examine whether candidate presentation order affects aggregate recommendation effectiveness, we conduct two-sided paired permutation tests on per-user NDCG@10 across all three permutation pairs for each decision-oriented model and candidate-set size. Each test uses 10,000 random sign-flip permutations, and Holm--Bonferroni correction is applied across 24 comparisons. None of the comparisons reaches statistical significance after correction ($p_{\mathrm{adj}} \geq 0.05$), indicating that we do not detect systematic differences in mean NDCG@10 across permutations.
Nevertheless, the substantial mean absolute NDCG@10 changes and Hit@10 flip rates reported in Table~\ref{tab:order_robustness} demonstrate that aggregate performance comparisons alone can conceal considerable variation in individual recommendation outcomes.

\subsubsection{Impact of Probability Ties}
\label{sec:tie}
To examine whether tied candidate probabilities influence our robustness findings, we additionally conduct a tie-aware evaluation that computes expected NDCG@10 and Hit@10 under uniformly randomized ordering within tied-score groups. As shown in Table~\ref{tab:order_robustness_tie}, the resulting mean absolute NDCG@10 changes remain close to those obtained using deterministic tie-breaking, with differences of at most 0.003 across all four candidate-set sizes in Movies and TV. Jev continues to exhibit lower overall ordering sensitivity than Decisions across all candidate-set sizes, although their informative-user sensitivity becomes comparable at larger candidate-set sizes. These findings suggest that the observed ordering sensitivity is not primarily driven by the deterministic resolution of probability ties.

\begin{table*}[t]
\centering
\caption{Tie-aware candidate-order robustness of Jev and Decisions on Movies and TV across candidate-set sizes. To account for tied candidate probabilities, we compute expected NDCG@10 and Hit@10 under uniformly random tie-breaking. For each user, we evaluate three permutations of the same candidate set (one original and two shuffled). Mean NDCG@10 is averaged across permutations, while Mean Absolute $\Delta$NDCG@10 and Mean Absolute $\Delta$Expected Hit@10 are averaged over all permutation pairs. Informative users are those achieving positive expected NDCG@10 under at least one permutation. The informative-user metrics are computed on this subset, with its size reported as $N$ Informative. Lower absolute changes and higher unchanged rates indicate greater robustness.}
\label{tab:order_robustness_tie}
\resizebox{\textwidth}{!}{
\begin{tabular}{lcccccccc}
\toprule
& \multicolumn{4}{c}{Jev} 
& \multicolumn{4}{c}{Decisions} \\
\cmidrule(lr){2-5}
\cmidrule(lr){6-9}
Metric 
& $K=20$ & $K=50$ & $K=100$ & $K=200$
& $K=20$ & $K=50$ & $K=100$ & $K=200$ \\
\midrule
Mean NDCG@10 & 0.295 & 0.165 & 0.113 & 0.078 & 0.270 & 0.138 & 0.089 & 0.056\\
Mean Absolute $\Delta$NDCG@10 & 0.139 & 0.077 & 0.054 & 0.028 & 0.173 & 0.097 & 0.059 & 0.043\\
Mean Absolute $\Delta$Expected Hit@10 & 0.267 & 0.159 & 0.102 & 0.049 & 0.335 & 0.184 & 0.100 & 0.076\\
Informative Mean Absolute $\Delta$NDCG@10 & 0.151 & 0.111 & 0.108 & 0.088 & 0.187 & 0.134 & 0.107 & 0.087\\
Informative-User Unchanged Rate & 0.106 & 0.099 & 0.093 & 0.110 & 0.058 & 0.033 & 0.038 & 0.017\\
N Informative & 877 & 665 & 474 & 300 & 882 & 691 & 525 & 464\\
\bottomrule
\end{tabular}}
\end{table*}

\subsection{RQ3: Efficiency and Scalability}
\label{appendix_sec:rq3}

\subsubsection{Latency Distribution} 
\label{sec:latency}
Table~\ref{tab:latency_distribution_stats} reports the median, 25th percentile, and 75th percentile of observed per-user inference latency across datasets and candidate-set sizes. Decisions consistently exhibits lower median latency than Jev, while both decision-oriented approaches are substantially faster than GPT-6-Sol and GPT-6-Luna. Their latency increases relatively moderately with candidate-set size, in contrast to the more pronounced increases observed for Qwen-based rerankers and GPT-based listwise reranking. However, the comparison between hosted and locally executed methods reflects their respective deployment environments rather than hardware-normalized inference efficiency.

\begin{table*}[t]
    \centering
    \caption{
Distribution of observed per-user inference latency across three recommendation domains and four candidate-set sizes. Each entry reports median [P25, P75] latency in milliseconds. (P) and (L) denote pointwise and listwise reranking, respectively. Qwen2.5 7B and 32B refer to the corresponding instruction-tuned models. Hosted API latencies include network communication and provider-side processing and are not hardware-normalized against locally executed methods.
    }
    \label{tab:latency_distribution_stats}
    \resizebox{0.65\textwidth}{!}{
    \begin{tabular}{lcccc}
        \toprule
        Method
        & $K=20$
        & $K=50$
        & $K=100$
        & $K=200$ \\
        \midrule
\multicolumn{5}{l}{\textbf{Movies and TV}} \\
SASRec & \lat[2]{1.165} {1.152} {1.183} & \lat[2]{1.205} {1.194} {1.223} & \lat[2]{1.184} {1.174} {1.200} & \lat[2]{1.189} {1.173} {1.216}\\
DCNv2 & \lat[2]{0.565} {0.556} {0.569} & \lat[2]{0.557} {0.552} {0.565} & \lat[2]{0.548} {0.544} {0.555} & \lat[2]{0.554} {0.549} {0.561}\\
Qwen2.5 7B (P) & \lat[2]{606.545} {524.313} {1305.220} & \lat[2]{1526.348} {1310.938} {3233.704} & \lat[2]{3044.906} {2597.272} {6449.878} & \lat[2]{6072.933} {5200.285} {12840.475}\\
Qwen2.5 7B (L) & \lat[2]{63.604} {61.785} {173.579} & \lat[2]{120.303} {116.140} {363.540} & \lat[2]{202.384} {196.555} {680.471} & \lat[2]{398.461} {382.391} {1413.075}\\
Qwen2.5 32B (P) & \lat[2]{2520.937} {2163.036} {5395.480} & \lat[2]{6252.235} {5365.620} {13302.156} & \lat[2]{12465.888} {10672.714} {26490.238} & \lat[2]{24925.080} {21354.697} {52958.626}\\
Qwen2.5 32B (L) & \lat[2]{280.465} {273.151} {740.674} & \lat[2]{501.373} {488.431} {1505.283} & \lat[2]{886.562} {852.869} {2808.216} & \lat[2]{1657.024} {1586.352} {5725.895}\\
GPT-6-Sol (L) & \lat[2]{1449.520} {1328.401} {1653.564} & \lat[2]{2722.062} {2494.141} {2978.314} & \lat[2]{4645.437} {4169.231} {5154.784} & \lat[2]{6777.843} {5705.748} {8158.964}\\
GPT-6-Luna (L) & \lat[2]{1319.666} {1198.396} {1484.090} & \lat[2]{1761.666} {1592.278} {1975.587} & \lat[2]{2369.761} {2109.095} {2806.398} & \lat[2]{3995.258} {3503.859} {6756.230}\\
Decisions & \lat[2]{91.589} {86.240} {99.425} & \lat[2]{136.185} {122.373} {148.430} & \lat[2]{146.050} {133.370} {164.663} & \lat[2]{160.843} {148.962} {186.895}\\
Jev & \lat[2]{154.573} {140.770} {169.663} & \lat[2]{159.387} {144.580} {174.487} & \lat[2]{179.230} {162.092} {205.172} & \lat[2]{198.178} {181.551} {260.242}\\
\midrule

\multicolumn{5}{l}{\textbf{Video Games}} \\
SASRec & \lat[2]{1.177} {1.167} {1.193} & \lat[2]{1.165} {1.155} {1.182} & \lat[2]{1.183} {1.172} {1.199} & \lat[2]{1.188} {1.178} {1.206}\\
DCNv2 & \lat[2]{0.550} {0.541} {0.556} & \lat[2]{0.547} {0.543} {0.555} & \lat[2]{0.563} {0.559} {0.570} & \lat[2]{0.553} {0.549} {0.565}\\
Qwen2.5 7B (P) & \lat[2]{1479.530} {1394.836} {1541.105} & \lat[2]{3645.464} {3463.944} {3812.011} & \lat[2]{7266.839} {6904.421} {7608.791} & \lat[2]{14555.192} {13850.509} {15193.401}\\
Qwen2.5 7B (L) & \lat[2]{181.398} {178.804} {188.337} & \lat[2]{368.393} {345.830} {374.243} & \lat[2]{687.969} {672.479} {700.006} & \lat[2]{1448.282} {1422.864} {1470.703}\\
Qwen2.5 32B (P) & \lat[2]{6062.837} {5797.198} {6393.015} & \lat[2]{15167.239} {14512.956} {15957.616} & \lat[2]{29905.051} {28458.954} {31333.674} & \lat[2]{60382.574} {57524.865} {63333.573}\\
Qwen2.5 32B (L) & \lat[2]{766.634} {755.685} {795.974} & \lat[2]{1551.064} {1506.270} {1567.628} & \lat[2]{2892.206} {2793.517} {2956.964} & \lat[2]{5798.239} {5654.639} {5899.738}\\
GPT-6-Sol (L) & \lat[2]{1552.489} {1428.686} {1740.031} & \lat[2]{2627.668} {2436.193} {2860.263} & \lat[2]{4471.988} {4169.283} {4785.484} & \lat[2]{9023.301} {8113.901} {10501.362}\\
GPT-6-Luna (L) & \lat[2]{1499.538} {1317.342} {1825.593} & \lat[2]{2388.700} {2123.335} {2859.586} & \lat[2]{3208.640} {2940.898} {3651.594} & \lat[2]{11972.634} {6995.724} {12931.694}\\
Decisions & \lat[2]{135.484} {125.460} {145.356} & \lat[2]{147.429} {138.339} {158.189} & \lat[2]{171.501} {161.611} {183.922} & \lat[2]{195.882} {183.093} {210.930}\\
Jev & \lat[2]{154.874} {145.248} {168.240} & \lat[2]{174.526} {165.555} {187.523} & \lat[2]{209.974} {200.175} {223.408} & \lat[2]{292.815} {272.108} {308.308}\\
\midrule

\multicolumn{5}{l}{\textbf{Books}} \\
SASRec & \lat[2]{1.161} {1.151} {1.176} & \lat[2]{1.137} {1.123} {1.154} & \lat[2]{1.200} {1.190} {1.218} & \lat[2]{1.158} {1.146} {1.173}\\
DCNv2 & \lat[2]{0.560} {0.551} {0.564} & \lat[2]{0.549} {0.544} {0.556} & \lat[2]{0.553} {0.547} {0.561} & \lat[2]{0.556} {0.547} {0.603}\\
Qwen2.5 7B (P) & \lat[2]{1524.835} {1395.821} {1617.516} & \lat[2]{3769.366} {3466.504} {3986.260} & \lat[2]{7494.929} {6908.141} {7929.719} & \lat[2]{14868.048} {13754.646} {15732.938}\\
Qwen2.5 7B (L) & \lat[2]{193.406} {186.986} {199.496} & \lat[2]{390.254} {380.122} {403.463} & \lat[2]{731.226} {709.083} {759.330} & \lat[2]{1530.139} {1482.892} {1584.416}\\
Qwen2.5 32B (P) & \lat[2]{6221.655} {5749.254} {6643.521} & \lat[2]{15565.381} {14351.194} {16600.549} & \lat[2]{31085.827} {28615.239} {33013.229} & \lat[2]{62314.826} {57517.370} {66338.040}\\
Qwen2.5 32B (L) & \lat[2]{799.306} {771.410} {853.536} & \lat[2]{1629.497} {1592.986} {1689.941} & \lat[2]{3096.692} {2979.131} {3168.128} & \lat[2]{6329.575} {6106.175} {6531.860}\\
GPT-6-Sol (L) & \lat[2]{1274.188} {1195.753} {1406.008} & \lat[2]{2020.629} {1891.557} {2175.181} & \lat[2]{3319.803} {3104.794} {3568.882} & \lat[2]{6383.789} {5891.093} {7211.060}\\
GPT-6-Luna (L) & \lat[2]{1399.218} {1252.379} {1597.523} & \lat[2]{2106.625} {1939.721} {2306.448} & \lat[2]{3470.816} {3089.163} {4196.430} & \lat[2]{12448.895} {7345.128} {14474.179}\\
Decisions & \lat[2]{137.009} {124.609} {148.451} & \lat[2]{148.134} {140.221} {159.169} & \lat[2]{170.349} {153.586} {182.417} & \lat[2]{192.744} {181.815} {204.091}\\
Jev & \lat[2]{152.284} {141.336} {168.539} & \lat[2]{163.412} {156.689} {175.229} & \lat[2]{202.084} {192.801} {213.386} & \lat[2]{264.068} {254.235} {277.848}\\

        \bottomrule
    \end{tabular}
    }
\end{table*}

\end{document}